\documentclass[a4paper,fleqn, nonatbib]{cas-dc}

\usepackage[natbibapa]{apacite}
\usepackage[utf8]{inputenc}
\usepackage{newunicodechar}
\newunicodechar{∗}{\textasteriskcentered}

\usepackage{threeparttable}
\usepackage{float}
\usepackage{booktabs}
\usepackage{url}
\usepackage{graphicx}
\usepackage{placeins}
\usepackage{xcolor}

\def\tsc#1{\csdef{#1}{\textsc{\lowercase{#1}}\xspace}}
\tsc{WGM}
\tsc{QE}
\tsc{EP}
\tsc{PMS}
\tsc{BEC}
\tsc{DE}

\begin{document}
\let\WriteBookmarks\relax
\def\floatpagepagefraction{1}
\def\textpagefraction{.001}

\shorttitle{How Students Really Use ChatGPT}

\shortauthors{Ammari et al.}

\title[mode = title]{How students (really) use ChatGPT: Uncovering experiences among undergraduate students}

\author[1]{Tawfiq Ammari}[
  orcid=0000-0002-1920-1625,
  type=editor,
  auid=000,
  bioid=1
]
\cormark[1]
\ead{tawfiq.ammari@rutgers.edu}
\affiliation[1]{organization={Rutgers University School of Communication and Information},
                addressline={4 Huntington Street},
                city={New Brunswick},
                state={NJ},
                postcode={08901},
                country={USA}}

\author[1]{Meilun Chen}
\ead{meilun.chen@rutgers.edu}

\author[1]{S M Mehedi Zaman}
\ead{sm.mehedi.zaman@rutgers.edu}

\author[1]{Kiran Garimella}
\ead{kg766@comminfo.rutgers.edu}

\cortext[1]{Corresponding author.}

\begin{abstract}
We examine how undergraduate students integrate ChatGPT into everyday self-directed learning. We analyze 10{,}536 naturalistic messages donated by 36 students over more than a year. A sequential mixed-methods pipeline pairs iterative team-based qualitative coding with zero-shot language-model annotation validated against human labels ($\kappa = 0.75$--$0.91$). It yields a five-category interaction taxonomy of Information Seeking, Content Generation, Language Use, Student--ChatGPT Interaction, and ChatGPT Response Behavior. Time-lagged linear regression and Cox proportional-hazards models link these categories to sustained engagement. Three findings stand out. First, structured, goal-driven tasks such as theory application, code writing, job-application content, and multiple-choice questions predict continued use, and ChatGPT becomes incorporated into students' academic rhythms when gratifications are reliably fulfilled. Second, system-issued \emph{apologies} are the single strongest positive predictor of increased engagement, outweighing every task-completion predictor. We name this mechanism \emph{repair gratification}, the reward of a breakdown acknowledged and repaired rather than a task simply completed. Third, behaviors marking interactional strain, including prompt revision, frustration, and follow-up clarification, predict disengagement. When the cost of managing the system falls on the user without system-side accountability, students abandon the tool. We interpret these results through a triangulation of Self-Directed Learning, Uses and Gratifications Theory, and Expectancy Violations Theory, mapping every predictor onto the four conditions of positive violation, positive confirmation, negative violation, and negative confirmation. We close with design recommendations for graduated repair patterns, mode-aware interaction, and verification affordances, and we outline a participatory AI-literacy agenda for higher education.
\end{abstract}

\begin{highlights}
\item Naturalistic logs from 36 undergraduates (10{,}536 messages; Dec 2022--Jan 2024).
\item Five-category interaction taxonomy, reliably coded by humans and a language model.
\item System apologies are the strongest positive predictor of returning the next week.
\item Repair gratification: reward from a breakdown repaired, not a task completed.
\item Every predictor of use and non-use maps onto expectancy violations.
\end{highlights}


\maketitle

\section{Introduction}
\label{sec:intro}

Large language models and AI-driven chatbots such as ChatGPT are reshaping how learners
search for information, plan assignments, and solicit feedback. We still lack a clear picture
of how students weave the tool into academic routines spanning formal coursework and the
informal, logistical practices through which they manage learning. Most of what we know comes
from self-reports, single-course case studies, or public prompt repositories. What students
actually ask ChatGPT to do, and how that use evolves across a semester, remain open questions
\citep{skjuve2024people, BEWERSDORFF2025100340, sawalha2024analyzing}.

This gap matters because institutions are already writing policies and building infrastructure
around generative AI \citep{jin2025generative}. OpenAI's May 2024 release of ChatGPT Edu,
developed with several U.S. universities, has intensified both investment and anxiety over
responsible deployment \citep{openai_chatgpt_edu, coffey_2024_chatgpt_highered}.
Administrators need evidence on whether the tool augments learning or erodes it,
instructors need guidance on scaffolding its use, and students need to understand the
trade-offs around trust, privacy, and academic integrity. Without a fine-grained account of
actual use, policies risk being either toothless or overly restrictive.
\citet{han2026ethical} show how unequal AI literacy among parents, administrators, and
developers weakens accountability for technologies whose burdens fall on students.

Studying this is methodologically hard. ChatGPT interactions are self-directed, cross
disciplinary boundaries, and change as the model and interface are updated. Existing work
relies on self-report surveys \citep{skjuve2024people, BEWERSDORFF2025100340,
sajawal2024design}, interviews \citep{shoufan, agrawal}, public datasets such as WildChat
\citep{zhao2024wildchat} or ShareGPT \citep{deng2023early}, and partial user logs
\citep{sawalha2024analyzing}, each a narrow slice that cannot track the iterative character of
real-world use. Laboratory prompts and single-session surveys miss what students do when
a response is confusing, hallucinated, or misaligned with a rubric, and prompting is itself a
skill refined over repeated revision \citep{cao2025study, Zamfirescu_et_al_23}. A single
interaction may serve academic, emotional, and logistical needs at once, resisting clean
functional labels. The literature is also fragmented by discipline, centering on computer
science and engineering \citep{hingle_25, uddin2024chatgpt, alves2024give,
mcnichols2025studychat, xue_et_al_24}, English as a Foreign Language writing
\citep{kim2024designing}, or isolated skills such as coding \citep{guner2025ai,
hartley2024artificial, haindl2024does, zonnchen2024impact, ghimire2024coding} and writing
\citep{levine2025students, woo2024teaching}. A holistic picture of how students engage with
ChatGPT across their broader learning workflows is still missing.

Against this backdrop we ask
\begin{quote}
\textbf{RQ1}. What categories of ChatGPT prompts do students employ on a daily basis?
\end{quote}

We also ask which learners remain invested as the semester unfolds. A student's first few
interactions with an online system foreshadow later outcomes, and log-based behavioral
features predict course completion, grade trajectories, and dropout risk
\citep{brooks_et_al_15, tyas2022students}. The underlying approach is not specific to
education. Time-lagged features drawn from interaction logs have been used to predict
outcomes across a range of domains, including voice-assistant satisfaction
\citep{kanungofirst}, sequential recommendation \citep{chen_et_al_22}, consumer search and
purchase \citep{ABDULLHADI2024e25034}, and residential energy demand \citep{lork_et_al_17}.

\begin{quote}
\textbf{RQ2a}. Which usage patterns correlate with continued or increased reliance on ChatGPT
over time?
\end{quote}

Disengagement is equally revealing. Survival models such as Cox proportional hazards have
traced abandonment for social media, medical sensors, and other technologies, uncovering
contextual factors that push users away \citep{fronza2011failure, schoenebeck_14,
battelino2015routine, kim_et_al_23}. Adapting them to educational AI lets us ask

\begin{quote}
\textbf{RQ2b}. Which usage patterns indicate a likelihood of continued, long-term use of
ChatGPT by students?
\end{quote}

Both temporal questions turn on what happens when an exchange goes wrong. Phrasing
effective prompts is difficult \citep{cao2025study, Zamfirescu_et_al_23, denny_et_al_24,
geroimenko2025key}, repairing user--ChatGPT conversations is harder still
\citep{braggaar2023breakdowns, Zamfirescu_et_al_23}, and earlier work
documents repair as a central source of friction in text-based agents
\citep{rapp_human_2021}. We therefore ask

\begin{quote}
\textbf{RQ3}. What are the main challenges of interacting with ChatGPT faced by students in
their daily use, and how do students resolve these challenges?
\end{quote}

\begin{figure*}
    \centering
    \includegraphics[width=1\linewidth]{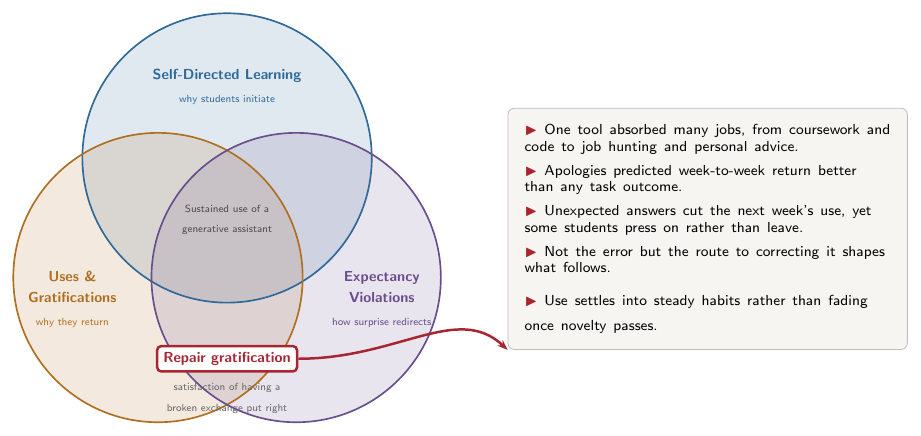}
    \caption{Conceptual overview. The three circles show the frameworks that jointly frame sustained use of a generative assistant, with the role each plays noted beneath its label.
Repair gratification, the construct our logs add, sits where gratification meets violation. The points at right summarize what the conversation logs contribute to that account.}
    \label{fig:triangulation}
\end{figure*}

We read these questions through three lenses, developed in \S\ref{sec:relwork} and shown
together in Figure~\ref{fig:triangulation}. \emph{Self-Directed Learning} frames how students
set goals and manage their own learning outside the classroom, and ChatGPT extends that setting
by offering on-demand, adaptive support \citep{song2007conceptual, li2024reconceptualizing,
baskara2023promises}. \emph{Uses and Gratifications Theory} explains what students seek from
the tool, whether utilitarian, hedonic, novelty-driven, or social, and what follows when those
expectations are disconfirmed \citep{Faruktechnicalgratification, jishnu2023unveiling,
liu2021ai}. \emph{Expectancy Violations Theory} provides the vocabulary for that breakdown-and-repair
dynamic, read here within the wider program of \emph{human--machine communication}, which
treats AI systems as communicative partners rather than tools
\citep{guzman_artificial_2020}. \citet{burgoon_application_2016} give EVT an operational form
by crossing expectancy with valence to yield four conditions, which we use as an analytic
device in \S\ref{sec:findings_evt} and intersect with gratifications in
\S\ref{sec:disc_repair}. Together the three lenses position ChatGPT as a socio-technical
partner embedded in learners' study ecologies rather than as a static tool.

Three contributions follow. First, a five-category taxonomy of student prompting derived
from donated conversation logs and validated against LLM annotation. Second, temporal models
separating what brings a student back the following week from what keeps them using the tool
across the term, which turn out to reward different behaviors. Third, and most consequential,
we find that ChatGPT's own \textbf{Apologies} are the strongest positive predictor of continued
use, outweighing every task-completion predictor in either model. We argue that this reflects a
gratification the existing UGT inventory does not name, which we call \emph{repair
gratification}, the return a student takes from a failed exchange being restored rather than
from a task being completed. Foreclosing that repair is equally consequential. \textbf{Stated Limitations}, in which the system declines before attempting the task, damage engagement at least as much as outright errors do, because it leaves the student with nothing to repair.

The paper is organized as follows. Section~\ref{sec:relwork} synthesizes related work
along the three triangulation legs. Sections~\ref{sec:dataset} and~\ref{sec:meth} describe the
donated-log dataset and the sequential mixed-methods pipeline. Section~\ref{sec:findings}
presents the taxonomy, the temporal models, and the expectancy-violation map of continued and
discontinued use. Section~\ref{sec:discussion} interprets these findings and develops repair
gratification. Sections~\ref{sec:design} and~\ref{sec:policy} translate the empirics into
design and policy implications, and Sections~\ref{sec:limitations} and~\ref{sec:conclusion}
acknowledge limitations and conclude.

\section{Related Work}
\label{sec:relwork}
We synthesize related work along the three legs that organize our analysis. We first situate
ChatGPT within \emph{Self-Directed Learning} and the intelligent learning environments that
support it. We then turn to \emph{Uses and Gratifications}, the mental models students bring
to an opaque system, and what happens when gratifications are disrupted. We close with
\emph{Expectancy Violations Theory} and the wider \emph{human--machine communication} agenda,
the closest precedents for our breakdown-and-repair findings.

\subsection{ChatGPT and Self-Directed Learning in Higher Education}
\label{sec:rw_sdl}
SDL is the learner-driven process of setting goals, identifying resources, applying
strategies, and evaluating outcomes \citep{knowles1975self}. Digital platforms such as
YouTube, MOOCs, and Duolingo have expanded it beyond formal education into informal,
technology-enabled learning \citep{onah2021innovative}, and advances in AI,
ChatGPT among them, have made it more dynamic and interactive \citep{curran2019adult,
li2024reconceptualizing}.

Interactive learning environments (ILEs) extend SDL by helping students plan, monitor, and
reflect \citep{collins1988computer, simonson2019teaching, reusser2013tutoring,
andres2023finding}. The affordances \citet{self1988student} and \citet{nwana1990intelligent} identify for
these environments (corrective, elaborative, strategic, diagnostic, predictive, and evaluative) position AI not as one more tool but as a contextual force shaping how
learners manage emotional, cognitive, and procedural demands.
\citet{dmello_gaze_2012}'s gaze-reactive \emph{Gaze Tutor} makes the affective side of
these affordances concrete, detecting boredom and redirecting attention so that
engagement becomes something sensed and repaired in flight rather than a fixed property
of the learner. We return to the affective trajectory in \S\ref{sec:rw_misalign}.

Extending \citet{song2007conceptual}'s model of personal attributes, self-regulation, and
context, \citet{li2024reconceptualizing} add AI-specific factors such as technology readiness
and socio-technical dynamics that shape the motivation to keep using a tool, a dynamic the
UGT framework below develops. ChatGPT promotes learner autonomy while raising concerns about
over-reliance, bias, and privacy \citep{li2024impact, lin2024exploring}. Our study
contributes large-scale naturalistic-log evidence on how, when, and for which task genres SDL
with a generative assistant persists, and on how such tools shift from novelty to
taken-for-granted parts of students' academic routines \citep{silverstone1992information,
haddon2007roger, bakardjieva2005internet}. Users fit a technology to their needs through
micro-regulations that settle into distinct \emph{use genres} \citep{bakardjieva2005internet},
a pattern traced for smart speakers \citep{Brause15042020}, energy-feedback displays
\citep{menon2025ambient}, and other ambient technologies \citep{marika_et_al_2010}. We use it here to characterize the use genres students settle into when directing
their own learning, and return to it when interpreting the survival results
(\S\ref{sec:disc_temporal}).

\subsection{Uses, Gratifications, and the Management of Misalignment}
\label{sec:rw_ugt}
We read students' motivations, and their responses to ChatGPT's limitations, through the Uses and
Gratifications Theory (UGT) lens \cite{katz1973uses}. We begin with the gratifications ChatGPT is known to deliver and
introduce \emph{repair gratification}, a further gratification that becomes available only
when an exchange fails. We then examine the mental models through which students interpret an
opaque system, and what follows when those models cannot account for what the system does. We
close with the work of repair itself, covering what students and systems do to close a
breakdown loop, the range of failures they have to close, and evidence that frustration does not
lead directly to abandonment.

\subsubsection{Uses and Gratifications}
\label{sec:rw_ug}
UGT explains why people turn to a technology by foregrounding the motivations that draw
them to it \citep{ruggiero2000uses, cho2017contextualization,katz1973uses}. Applied to students, it has
accounted for the pervasive uptake of networked platforms in university populations
\citep{IFINEDO2016192}. ChatGPT can deliver four gratifications for students.
\emph{Utilitarian} gratification is efficiency at academic tasks such as writing and coding
\citep{BRACHTEN2021102375}. \emph{Hedonic} gratification is enjoyment in interacting with a
conversational agent \citep{jo2022continuance, ammari_et_al_19}. \emph{Social} gratification
covers interaction or perceived companionship, often parasocial
\citep{xie2023understanding, maeda2024human}. Finally,  \emph{Creative} gratification is the sense of
amplified co-production students report when the system contributes to their work
\citep{lin2024unraveling}.

These four have something in common. Each is obtained from what the system delivers, whether
a finished task, an enjoyable exchange, a sense of company, or a co-produced artifact. We
introduce a further gratification that operates at a different level. \emph{Repair
gratification} is obtained not from what the system delivers but from how it conducts itself
once delivery has failed. It becomes available only when an exchange breaks down, and it is
earned when the system acknowledges the breakdown and works with the student to close it. A
student who receives a wrong answer, says so, and is met with an acknowledgment and a
correction has gained something a student whose first answer was simply right never had
access to.

The Computers-Are-Social-Actors (CASA) paradigm explains why a repaired breakdown should
register as a gratification rather than as damage merely undone. CASA holds that people
apply social rules to computers without deciding to \citep{nass1994computers,
reeves1996media, nass2001does, nass2000machines}, and acknowledging fault and making amends is among the
more strongly scripted of those rules \citep{Ashktorab_2019, zeng2024chatbot}. A system
that performs it is doing recognizably social work, and the student responds in kind. \citet{lee_enhancing_2017} give
the mechanism its most direct support, showing that a conversational agent's
\emph{reciprocity}, the return of effort or information to the user, predicts continued use
more strongly than self-disclosure does. What sustains use is the system's responsiveness to
the student's own turn rather than anything it volunteers unprompted, and repair is that
responsiveness under the least favorable conditions. We develop the construct in
\S\ref{sec:disc_repair}, where we show that it layers on the other gratifications rather than
replacing them. Repair is not equally available to every student, since a breakdown has to be
noticed and named before it can be closed. The capacity for repair therefore depends on the
user's mental model of the system.

\subsubsection{Mental Models of ChatGPT}
\label{sec:rw_mental}
Users interpret opaque systems through simplified mental models or folk theories
\citep{johnson1983mental, gelman2011concepts, devito2017algorithms, kelly_et_al_23}.
Undergraduate students, for instance, picture the internet as a ``huge information resource'' with no
sense of the underlying infrastructure \citep{zhang2008undergraduate}. When these models
misalign with how a technology works, confusion follows, especially for newer tools whose
limits users have not yet learned \citep{collings2002developing, druce2021brittle,
desai_et_al_metaphor_23} and when the tool is used to teach complex concepts
\citep{hamerska2024between, riedl2019human, jena2020c}.

Recent work sharpens this for ChatGPT. \citet{rapp_how_2025} show that nonsensical
hallucinations interact with prior expertise. Users with a computer-science background or
heavy LLM use read an unpredictable response as an \emph{error}, return it to a machine-frame,
and continue, whereas novices read it as the model's \emph{autonomous} act, which produces
uncanniness and generalized fear of AI. The breakdown is not the response but the absence of a
model that can absorb it, a difference visible in our own data, where students vary widely in
the load they bear when ChatGPT misbehaves.

\subsubsection{Misalignment, Breakdown, and Repair}
\label{sec:rw_misalign}
Users of unfamiliar tools reduce ambiguity and build trust by seeking cues and feedback
\citep{liu2021ai}, and their iterative experimentation functions as a repair mechanism when an
exchange breaks down \citep{liu2021ai, aquilino2024trust, thirunavukarasu2023large,
park_ahn_24}. Repair work is also available on the system side. \citet{Ashktorab_2019} find
that presenting the user with a set of options is among the most helpful strategies a system
can offer, even though doing so adds clutter to the interface.

What has to be repaired is varied. \citet{kim_et_al_24} catalog a wide range of
dissatisfactions that users report with conversational AI, and the wider literature documents
each of them. Systems misunderstand user intent \citep{RAO2023990} and adopt a mismatched tone
\citep{RAY2023121}. They refuse requests obstructively \citep{bang2023multitask, guo2023close,
borji2023categorical} and return content that is vague or unoriginal
\citep{borji2023categorical, kitamura2023chatgpt, kumar2023analysis}. They state facts
incorrectly or draw on stale knowledge \citep{azaria2022chatgpt, yeo2023assessing,
park_ahn_24}, contradict themselves across turns \citep{alkhamissi2022review}, reason
illogically \citep{zhang2023one}, hallucinate content outright \citep{ji2023survey}, and
defer to users even when those users are wrong, a tendency known as sycophancy
\citep{perez2023discovering}.

Work on affective trajectories complicates any simple story in which frustration leads
directly to abandonment. \citet{baker_better_2010} show that frustration is both less
persistent and less damaging to learning than boredom is, and \citet{kapoor_automatic_2007} show that frustration can be predicted from multimodal cues,
and multimodal sensing of learners' affective and cognitive states has since become a
working method inside live learning environments
\citep{cosentino_exploring_2025}. Frustration is therefore something a system could act on
rather than merely absorb. The design question is
therefore not whether breakdowns occur, since they will, but whether the system and the
learner close the repair loop before frustration hardens into disengagement. We return to this in \S\ref{sec:findings_temporal}.

\subsection{Expectancy Violations and Human--Machine Communication}
\label{sec:rw_evt}
SDL frames what students do and UGT why they engage. Our third theoretical leg explains the relational
mechanics of conversing with a generative AI, covering what users expect, how those
expectations are violated in both directions, and which violations sustain or break the
relationship. We anchor it in Expectancy Violations Theory (EVT), within the broader
program of human--machine communication (HMC).

\subsubsection{From HCI to Human--Machine Communication}
\label{sec:rw_hmc}
\citet{guzman_artificial_2020} argue that AI no longer fits the human--human model that long
organized communication research, and propose an HMC agenda that treats the machine as a
communicative entity. They organize it around three concerns, the \emph{functional} question
of how users make sense of AI as a communicator, the \emph{relational} question of how users
associate with these systems, and the \emph{metaphysical} blurring of human, machine, and
communication. The functional question is where chatbots most visibly strain. These systems
simulate human conversation but fall short on nuance, producing frustration and distrust
\citep{luger2016thegulf, brandtzaeg2018chatbots}, and users cope either by rephrasing their
prompts or, through automation bias, by accepting wrong answers on the strength of prior good
experiences \citep{li2020conversation, zaroukian2017automation}. The relational question
carries its own dynamics. In a two-wave study of AI voice assistants, \citet{xu_tool_2024}
show that functional and relational use each reinforce themselves over time but cross-couple
only loosely, with relational use through self-disclosure raising later functional use while
functional use alone tends to \emph{reduce} subsequent relational engagement. Students who
treat ChatGPT only as a query box may therefore follow a different long-run trajectory from
those who fold relational moves in, even when both report the same task on a given day.

\citet{pan_humanai_2025} add an encounter-level account. Studying initial human--AI exchanges,
they find that AI framed as more agentic, through an autonomous source and subjective
language, is rated more trustworthy and likable, most strongly when both signals occur
together and with users' AI literacy moderating the effect. What a user brings to first
contact therefore shapes what follows, and the affective channel is part of that. Confusion
can drive abandonment whereas empathy raises engagement \citep{fitzpatrick2017delivering}, and
parasocial attachment and anthropomorphic cues calibrate trust alongside perceived reliability
\citep{maeda2024human}. This is the first-encounter analog of the asymmetry
\citet{rapp_how_2025} document in extended use, since how a user reads the system on first
contact, and how richly they can model it, sets the slope for later trust and repair.

\subsubsection{Engagement and Relationship Formation in Text-Based Chatbots}
\label{sec:rw_chatbots}
Research on text-based chatbots is our closest empirical precedent, and across it
engagement rests on more than task completion. \citet{rapp_human_2021} review a decade of
the field and find satisfaction, trust, emotional involvement, and humanness recurring
alongside task outcomes. \citet{skjuve_my_2021} trace that relational element over time,
proposing a three-stage arc from \textit{explorative} to \textit{affective} to
\textit{stable} engagement among long-term Replika users.
\citet{haugeland_understanding_2022} locate it in conversational structure, showing that
topic-led customer-service exchanges strengthen anthropomorphism and hedonic quality
relative to task-led ones at some cost to efficiency.

Both of the later findings transfer to our setting in modified form. The task-led and
topic-led distinction maps onto our taxonomy, where \emph{Information Seeking} and
\emph{Content Generation} are task-led and \emph{Student--ChatGPT Interaction} is topic-led,
and the two call for different design support. The staged arc is partly visible, since
students skip orientation, bring breadth and depth almost immediately, and re-engage where
they perceive nonjudgment. Whether companion-chatbot dynamics transfer to a tool positioned
as informational and academic is an open question, and our findings on apologies as repair
suggest that they partly do. We extend this record with a large naturalistic-log study of
which interaction categories sustain engagement once novelty has faded.

\subsubsection{Expectancy Violations and Repair as a Positive Violation}
\label{sec:rw_evt_repair}
\citet{burgoon_application_2016} anchor the EVT leg in this journal, extending the theory
from interpersonal encounters to interaction with a machine partner. Users carry
expectations into any exchange \citep{burgoon1978communication}. Those expectations can be
confirmed or violated, and the act itself is appraised as favorable or unfavorable
\citep{burgoon1993interpersonal}. Crossing expectancy with valence yields four conditions.
A \emph{positive violation} is unexpected and favorable, a \emph{positive confirmation}
expected and favorable, a \emph{negative confirmation} expected and unfavorable, and a
\emph{negative violation} unexpected and unfavorable. Counter to intuition, violations can
beat confirmations \citep{burgoon1988nonverbal}, because departing from expectation draws
attention and makes valence more salient, so a favorable surprise moves relational
judgments further than a favorable routine.

Three of their results bear on ours. The positive-violation condition produced the highest task
attractiveness of any partner, humans included. Positively valenced conditions elicited more
connectedness, understanding, and dependability than negatively valenced ones. And, against
prediction, negative violations were no worse than negative confirmations.
\citet{burgoon_application_2016} caution that merely meeting expectations does not guarantee
good outcomes, and argue that agents should be built to evoke positive violations.

These foreshadow our central finding. A system that visibly acknowledges and repairs error does
something users do not expect a chatbot to do well, so it reads as a positive violation and as
a form of the reciprocity \citet{lee_enhancing_2017} identify, and it sustains engagement. This is why we treat \emph{repair} rather than \emph{accuracy} as the headline mechanism. An
error opens a gap between what the student expected and what the system did, and the
apology is the system's bid to repair the relationship that gap disturbs. EVT gives the
reason to watch that move rather than the correction itself, since departures from
expectation are weighted more heavily than routine success.

\citet{burgoon_application_2016} report a null that gives a second expectation we can test.
If ChatGPT's unexpected errors and its anticipated refusals are both unfavorable, the two
should prove similarly costly rather than the surprise proving worse.
\S\ref{sec:findings_evt} compares our coefficients against that expectation.

\section{Dataset}
\label{sec:dataset}

The corpus analyzed here consists of ChatGPT chat histories donated by 36 undergraduates enrolled at a public research university in the U.S. Northeast. Recruitment ran in two waves off flyers posted around campus, with 12 students enrolling in October 2023 and a further 24 in January 2024, and each completed donation was compensated at \$10.

Each donation is a full account export retrieved from \url{chat.openai.com} and uploaded through a secure study site, so what we analyze is the whole of a student's recorded history with the system rather than a self-selected excerpt of it. Exports of this kind carry no direct identifiers, but the substance of what students type can still be revealing, and we handled the material accordingly. Identifying details were stripped during cleaning, no demographic metadata were gathered, and the donation pipeline retains no mapping from any log back to the student who supplied it.

\begin{figure*}[H]
    \centering
    \includegraphics[width=0.95\linewidth]{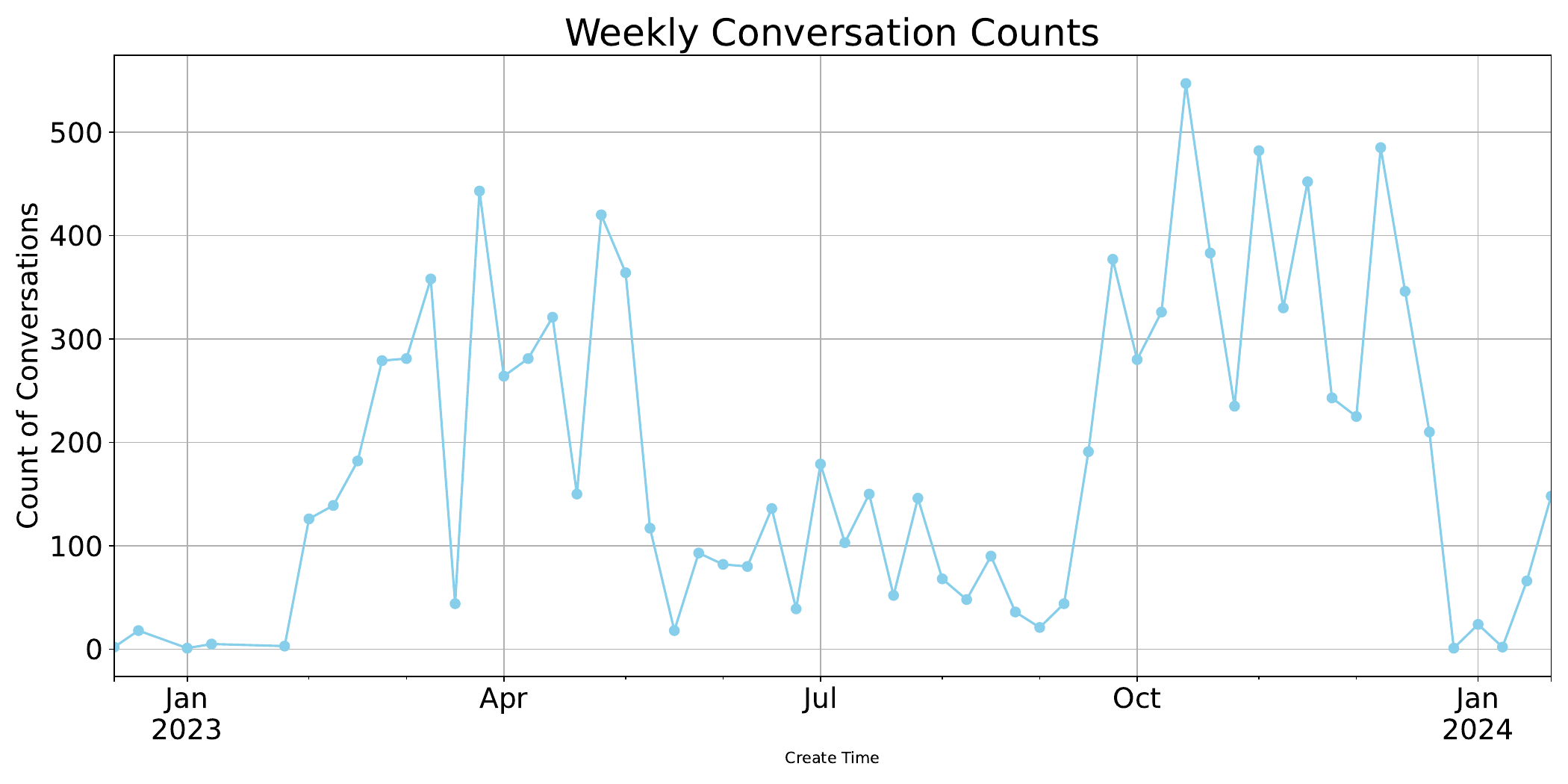}
    \caption{Conversation volume over time. The troughs coincide with Spring break (March 9--17) and Summer 2023, consistent with use that is largely tied to coursework.}
    \label{fig:timeseries}
\end{figure*}

Every row of the resulting file carries an internal user identifier unrelated to any real student ID, a conversation title the system generated automatically, a conversation identifier, a creation timestamp, the student's text, and the model's reply. Table~\ref{tab:dataset} reports the headline figures. Coverage runs a little over a year, from December 2022 to January 2024, and students averaged 45 sessions apiece (SD~$=$~66) at a mean duration of 13.2 minutes. Because each message is timestamped and grouped by student, the file sustains the week-level modeling described in \S\ref{sec:meth} rather than cross-sectional description alone. Figure~\ref{fig:timeseries} plots conversation volume over time. The rhythm tracks the university calendar, thinning over Spring break in March, across summer 2023, and around the December holidays.

\begin{table}[htbp]
\centering
\caption{The corpus at a glance.}
\label{tab:dataset}
\begin{tabular}{lr}
\toprule
\textbf{Characteristic} & \textbf{Value} \\
\midrule
Users & 36 \\
Unique chats & 1{,}631 \\
Messages & 10{,}536 \\
Mean session duration & 13.2 min \\
Period of coverage & Dec 2022 -- Jan 2024 \\
\bottomrule
\end{tabular}
\end{table}

 \section{Methods}
\label{sec:meth}
\begin{figure*}[H]
    \centering
    \includegraphics[width=1\linewidth]{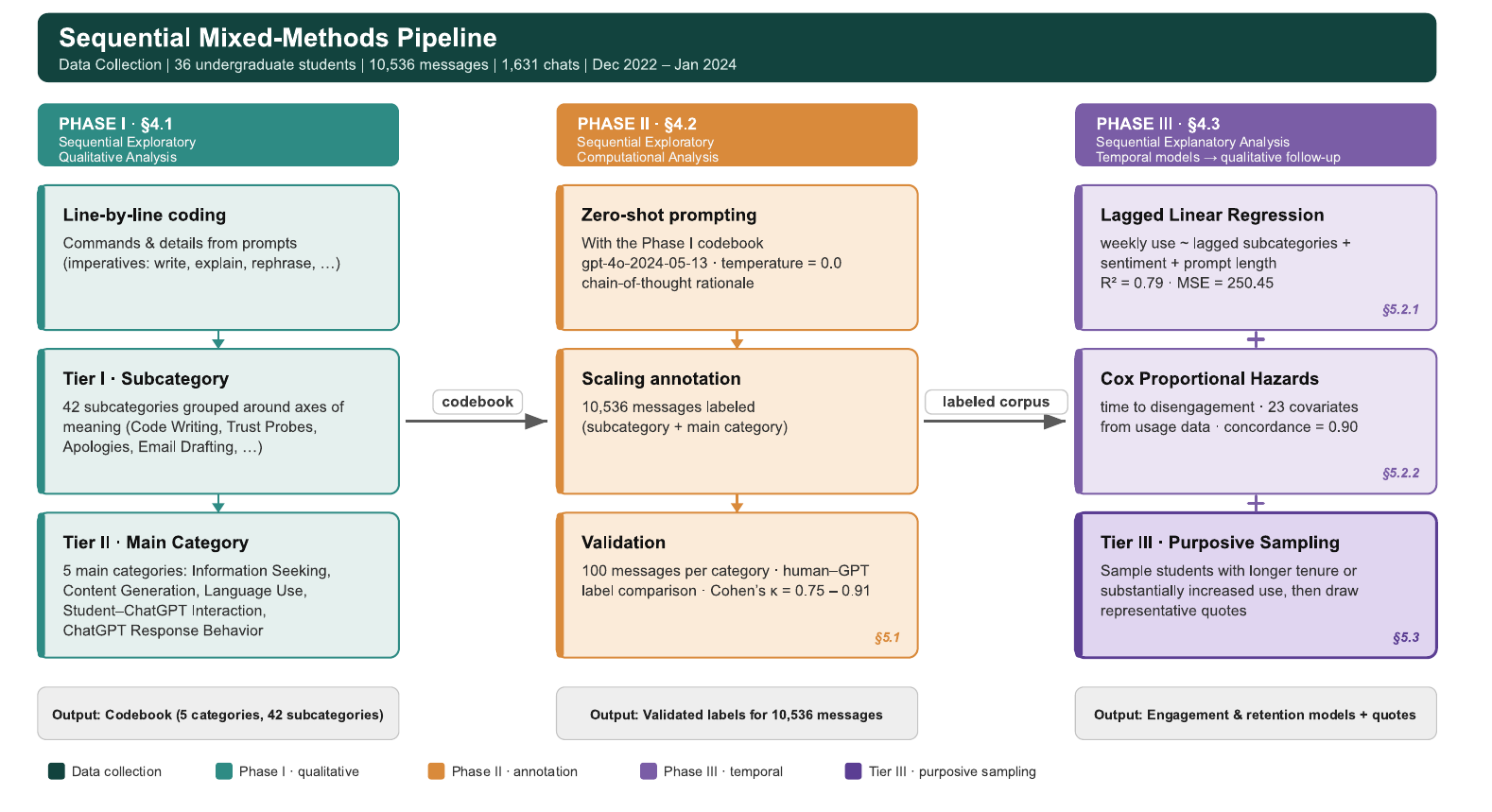}
    \caption{Sequential mixed-methods pipeline. Human researchers developed a grounded-theory codebook (Phase~I); this structured GPT-4o annotation across 10{,}536 messages (Phase~II); validation confirmed strong human--machine agreement ($\kappa = 0.75$--$0.91$), enabling temporal analyses linking interaction patterns to engagement trajectories (Phase~III).}
    \label{fig:meth_sum}
\end{figure*}

Our analysis proceeds in three phases with two different mixed-methods logics. Phases~I and~II follow a sequential exploratory strategy in which a qualitative stage builds an instrument that a subsequent computational stage applies at scale. Phase~III inverts that order and follows a sequential explanatory strategy in which quantitative results select the cases for a closing qualitative pass \citep[pp.~193--195]{creswell2017research}. Phase~I develops a codebook from manual coding of a message sample (\S\ref{sec:qualitative_coding}). Phase~II applies that codebook to the full corpus with an LLM annotator and validates the output against human judgment (\S\ref{sec:gpt_label}). Phase~III treats the validated labels as a longitudinal measurement instrument, models how the composition of a student's week relates to their later use, and then returns to the logs of the students the models single out (\S\ref{sec:phaseIItemp}). Figure~\ref{fig:meth_sum} summarizes the pipeline.

Because the coded categories become model covariates rather than an interpretive endpoint, we number the coding hierarchy from the message upward. Line-by-line annotation stays close to the wording of the prompt. \emph{Tier~I} is the subcategory at which coders converge. \emph{Tier~II} is the main category into which subcategories roll up. \emph{Tier~III} is not a coding level but a sampling pass keyed to a student's engagement trajectory, introduced in Phase~III once the temporal models identify which trajectories warrant closer reading. Main categories and their subcategories are defined in Appendix~\ref{sec:appendixA}. Appendix~\ref{sec:appendixB} documents the coding procedure, the construction of model covariates, and the Tier~III sampling thresholds.

\subsection{Phase I: Building the Coding Instrument}
\label{sec:qualitative_coding}
Phase~I produced a set of categories stable enough to be counted week by week and applied consistently by a machine annotator to messages no human had read. That requirement shaped how we coded. We needed categories whose boundaries could be stated operationally, whose membership did not depend on the surrounding conversation, and whose granularity was fine enough to distinguish interactional moves that we expected to behave differently over time, such as a student correcting the system and a student thanking it.

Two researchers coded a random sample of 1{,}882 messages using iterative-inductive thematic analysis as operationalized for team-based coding by \citet{cascio2019team}, drawing on the modified grounded-theory framework of \citet{corbin1990grounded}. Coders worked from the surface of each prompt outward, first recording what the student asked for in the student's own terms and then grouping those annotations around axes of meaning into 42 Tier~I subcategories. The 42 subcategories roll up into five Tier~II main categories that organize the analysis, namely \textit{Information Seeking}, \textit{Content Generation}, \textit{Language Use}, \textit{Student--ChatGPT Interaction}, and \textit{ChatGPT Response Behavior}.

Two features of the procedure matter for how the later models should be read. First, our stance was one of intercoder consensus rather than interrater reliability \citep{cascio2019team}. Coders worked as a single analytic team and treated an initial disagreement as evidence that a category boundary was underspecified, so disagreements were resolved by revising the definition rather than by averaging judgments or referring the case to a third coder \citep{mcdonald_reliability_2019}. Convergence at Tier~I was the operative signal of agreement where idiosyncratic line-by-line wording was expected. Second, each message received exactly one Tier~I label. A prompt that both requests code and expresses frustration was resolved to whichever move the prompt was doing, not split across categories. This keeps the weekly feature vectors of \S\ref{sec:phaseIItemp} interpretable as a composition of a student's activity, at the cost of flattening prompts that carry more than one move. Coding continued until no new subcategories emerged across successive transcripts, indicating saturation \citep[pp.~193--194]{creswell2017research}. The consensus procedure, the line-by-line mechanics, and worked examples tracing individual prompts through the hierarchy appear in Appendix~\ref{sec:appB-consensus} through Appendix~\ref{sec:appB-worked}.

\subsection{Phase II: Scaling Annotation to the Full Corpus}
\label{sec:gpt_label}
Given the dataset's size, we scaled analysis using GPT-based annotation \citep{chae_davidson_25}. We employed OpenAI's GPT-4o model (version: gpt-4o-2024-05-13, accessed July 2024) with temperature set to 0.0. We used a zero-shot prompt built from the human-developed codebook:

\begin{quote}\small\ttfamily
``The following are codes for qualitative analysis. You need to categorize the texts in inputs as one code from the following list and also reply with your chain of thought (COT) for the selection. Reply with **only the label in single quotation**, and then include COT in the next line without quotation. Remember, you are bound by only 1 label to reply.''
\end{quote}

For each message, the model returned a top-level category and a chain-of-thought rationale, supporting transparency and interpretability \citep{wei2022chain}. To assess reliability, we compared GPT and human labels for 100 messages per category. Agreement was substantial (Cohen's $\kappa$ = 0.75--0.91) \citep{mchugh2012interrater}, providing confidence in our automated annotation (Table~\ref{tab:kappa_scores}).

Two properties of this design bound what the annotation can support. The single-label constraint carries over from Phase~I, so corpus counts are counts of a message's dominant move rather than of everything present in it. We present the excerpts of Appendix~\ref{sec:appendixC} as the evidentiary anchor for claims about what repair sequences look like.

\begin{table}[ht]
\centering
\caption{Cohen's Kappa scores across annotation categories.}
\begin{tabular}{lc}
\toprule
\textbf{Category} & \textbf{Kappa} \\
\midrule
Information Seeking & 0.81 \\
Content Generation & 0.85 \\
Student--ChatGPT Interaction & 0.83\\
ChatGPT Response Behavior & 0.75 \\
Language Use & 0.91 \\
\bottomrule
\end{tabular}
\label{tab:kappa_scores}
\end{table}

\subsection{Phase III: Modeling Engagement over Time}
\label{sec:phaseIItemp}
Phase~III converts the labeled corpus into a panel. The observation unit is the student-week, chosen because the corpus is bounded by the academic calendar and because a week is the shortest interval over which most students in the sample generate enough messages for a stable composition. For each student-week we recorded the count of messages falling in each of the 42 subcategories derived in \S\ref{sec:qualitative_coding}, together with two controls, namely the composite valence score from VADER \citep{hutto2014vader} and mean prompt length. Valence is included because affect is confounded with several interactional subcategories, and length because longer prompts carry more task specification and would otherwise load onto the content-generation covariates. Predictors were standardized so that coefficients are comparable across covariates with very different base rates.

\subsubsection{Effect of Recent Use History on Future Use}
We defined weekly session length as the number of interactions per student per week. To account for temporal dynamics, we created lagged versions of each category, reflecting their values in the previous week, following lagged-regression precedents from electronic-health-record analysis \citep{levine_methodological_2018,levine2017comparing} as implemented in \citep{cubed2023lagged, scikit2024lagged}. Only students with at least two weeks of data were included after applying \texttt{.groupby().shift(1)} to compute lags. The final model was a linear regression, using the lagged feature set as independent variables and the following week's session length as the dependent variable; predictors were standardized with StandardScaler. The model achieved $R^2 = 0.79$ and MSE $=$ 250.45. The one-week lag means every reported association runs from a completed week to the week after it, which rules out same-week reverse causation but not confounding by a slower-moving cause such as a heavy assignment period.

\subsubsection{Survival Analysis}
We used a Cox Proportional Hazards (PH) model \citep{cox1972regression} to examine how usage behaviors predict time to disengagement. Implemented via \texttt{lifelines}'s \texttt{CoxPHFitter} \citep{Davidson_Pilon2019}, the model used time-to-disengagement (in weeks) and a binary disengagement flag. All 36 observations were uncensored. After removing collinear variables, the final model included 23 covariates from coded usage data; baseline hazards were estimated using the Breslow method. The model achieved a concordance index of 0.90, a partial log-likelihood of $-67.97$, and a log-likelihood ratio test of 41.21 (23 df, $-\log_2(p) = 6.48$).

Two features of the design constrain interpretation. Disengagement is observed as the end of a student's donated history, so what the model estimates is time to the last message a student contributed rather than time to abandonment of ChatGPT as such. A student who continued using the system after donating would appear here as disengaged. Relatedly, the absence of censoring is a property of how the data were collected rather than a finding, since every history terminates at the donation date. Hazard ratios should therefore be read as describing the shape of the donated record and not as forecasts beyond it.

\subsubsection{Tier III: Purposive Sampling by Engagement Trajectory}
\label{sec:tier3}

Tiers~I and~II characterize what students did with ChatGPT and how those behaviors
group, but neither is keyed to how a student's use developed over time. The temporal models
identify which behaviors predict two distinct outcomes, a longer span of continued use and
a rising volume of use from one week to the next, yet they do not identify what those
behaviors felt like to the student. Tier~III addresses both gaps by letting the
quantitative results guide a qualitative follow-up, following the sequential explanatory
logic in which a qualitative stage builds on and helps explain an initial quantitative one
\citep[pp.~193--194]{creswell2017research}.

Sampling was keyed to the two outcomes the models predict. \emph{Tenure}, the span in
weeks between a student's first and last donated message, is the trajectory counterpart of
the survival results, and \emph{intensification}, the ratio of message volume in the second
half of a student's active span to the first, is the counterpart of the weekly regression.
Tenure in the top tertile ($\geq 35$~weeks) was treated as sustained use and a ratio of at
least $1.5$ as intensifying use, with 21 students meeting at least one criterion. Drawing
illustrative material from these students rather than from the corpus at large keeps the
qualitative tier anchored to the sustained and intensifying use the temporal models
foreground rather than to one-off exchanges. Thresholds are detailed in
Appendix~\ref{sec:appB-tier3}.

Within these sampled logs we addressed RQ3 by placing interactional subcategories in
the two-dimensional space Expectancy Violations Theory specifies, crossing whether an
interaction confirmed or violated what the student appeared to expect with whether its
valence was positive or negative \citep{burgoon_application_2016}. Assignment was
interpretive and made at the level of the exchange rather than the message, since a
violation is legible only against the expectation the preceding turns establish. Expectancy
and valence were inferred from the student's next turn rather than measured directly, a
limitation we return to in \S\ref{sec:limitations}. Two researchers independently placed
each sampled exchange in a quadrant and reconciled differences through discussion, following
the consensus principle used in Phase~I. Where a session contained violations in more than
one direction we recorded them separately rather than assigning the session a net valence.
The resulting map appears in \S\ref{sec:findings_evt} and the underlying exchanges in
Appendix~\ref{sec:appendixC}.

\section{Findings}
\label{sec:findings}
 
We present our findings in four parts. Section~\ref{sec:findings_taxonomy} presents the five-category interaction taxonomy that answers RQ1 and gives empirical shape to RQ3. Sections~\ref{sec:findings_regression} and~\ref{sec:findings_survival} present the time-lagged regression and Cox survival models that answer RQ2 and RQ3. Section~\ref{sec:findings_evt} then maps every significant predictor onto the four EVT conditions of \citet{burgoon_application_2016}, which lets us say not only which behaviors predict continued use but why.
 
\subsection{A Five-Category Interaction Taxonomy (RQ1)}
\label{sec:findings_taxonomy}
We organize our findings around five themes, Content Generation, Information Seeking, Language Use, Student--ChatGPT Interaction, and ChatGPT Response Behavior. The first three reflect functional use; the latter two capture relational and system-side dynamics. The overall breakdown is presented in Figure~\ref{fig:complete}; sub-categories and their distributions are detailed in Appendix~\ref{sec:appendixA}.
 
\begin{enumerate}
\item \textbf{Information Seeking}: This category involves retrieving factual information, clarifying concepts, or answering specific questions. Students sought help on academic topics, job applications, medical issues, and social or cultural matters, reflecting everyday information needs \citep{savolainen1995everyday}. Example commands include:
\begin{quote}\small\ttfamily\noindent\strut{}``difference between disinformation and missinformation''; ``how to find key information on youtube''; ``are tuples mutable''
\end{quote}
 
\item \textbf{Content Generation}: This category covers the creation of original content such as essays, code, poems, emails, and resumes. Prompts varied across academic subjects, job applications, and other tasks emphasizing creativity and stylistic input.
\begin{quote}\small\ttfamily\noindent\strut{}``write a bullet point about me using agile software development cycle at an internship for my resume''; ``why mgiht a function prototype be necessary?''
\end{quote}
 
\item \textbf{Student--ChatGPT Interaction}: This category captures socially meaningful engagement with ChatGPT, follow-up questions, emotional responses, and signs of treating the AI as a conversational partner. A rare subset (1\%) included role-play and fictional dialogues.
\begin{quote}\small\ttfamily\noindent\strut{}``You're wrong, fix it immediately''; ``its actually right tho ur wrong, so when I ask the question again say the correct answer and why''
\end{quote}
 
\item \textbf{ChatGPT Response Behavior}: This category focuses on how ChatGPT responds to students, especially in problematic or unsatisfying interactions \citep{rapp_human_2021}, when it misinterprets prompts, adjusts its response style and complexity, or reports a limitation to the user.
\begin{quote}\small\ttfamily\noindent\strut{}``I don't have access to real-time information or current events beyond my last training data in September 2021''; ``You're right, apologies for the confusion. There are indeed three cases to consider when removing a node from a circular linked list.''
\end{quote}
 
\item \textbf{Language Use}: This category includes paraphrasing, synonym/antonym suggestions, rhetorical-style adjustments, grammar checks, and translation, tasks where students provided text and ChatGPT refined it.
\begin{quote}\small\ttfamily\noindent\strut{}``reword this to add details that the \$11,000 was used for purchasing food, decorations, tent rentals, and hiring help''; ``write this more professionally''
\end{quote}
\end{enumerate}
 
\begin{figure*}[H]
    \centering
    \includegraphics[width=1\linewidth]{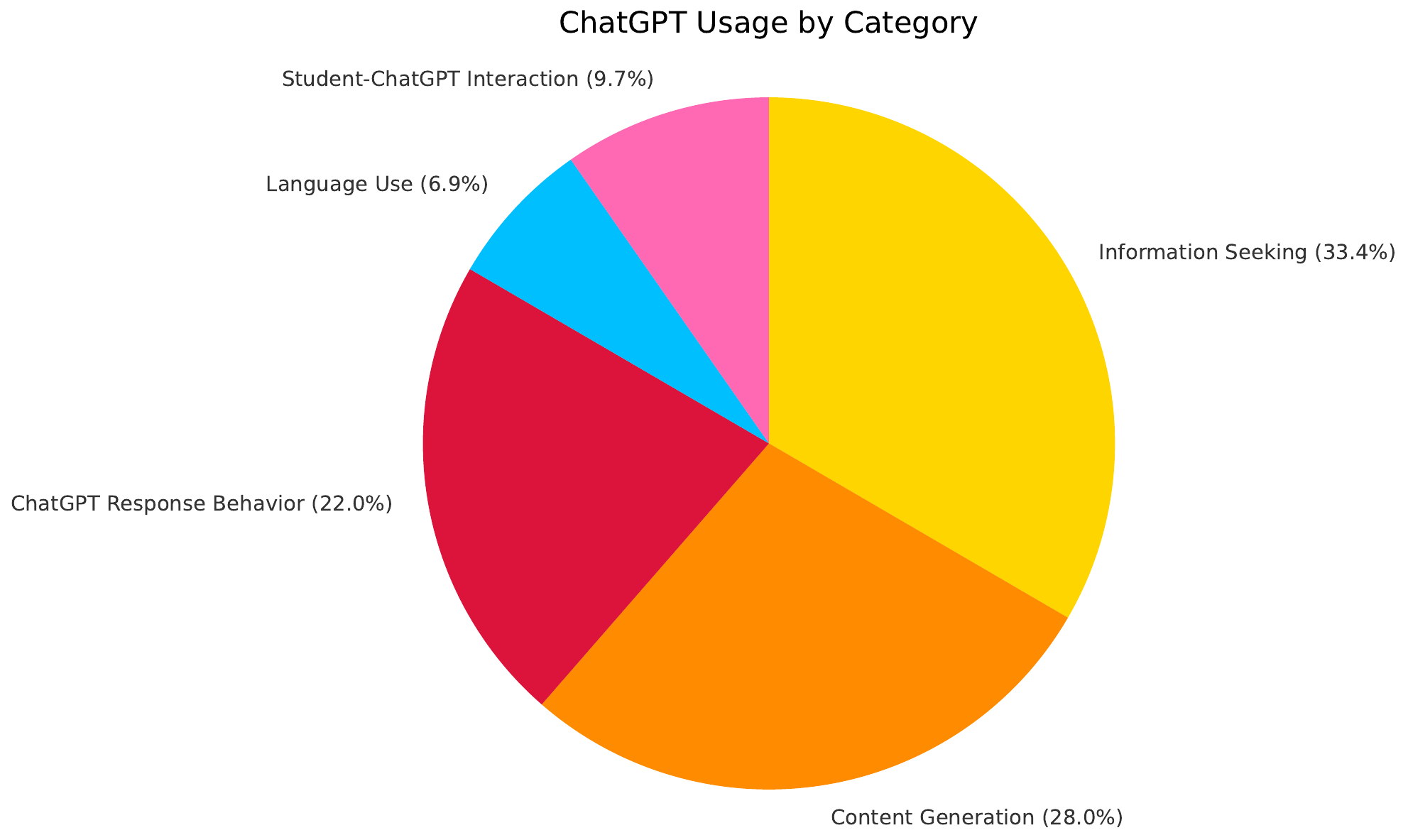}
    \caption{Breakdown of log interactions across the five top categories.}
    \label{fig:complete}
\end{figure*}

\subsection{Predicting Increased Use and Sustained Engagement (RQ2)}
\label{sec:findings_temporal}
We model continued use at two timescales. RQ2a (\S\ref{sec:findings_regression}) asks how \emph{much} a student's use of ChatGPT increases week over week while RQ2b (\S\ref{sec:findings_survival}) asks how \emph{long} they keep using it before dropping out.
 
\subsubsection{Predictors of Increased Use (RQ2a)}
\label{sec:findings_regression}
To test how prior behavior shapes future engagement, we regressed each student's weekly volume on the previous week's subcategory counts, standardized and lagged by one week (\S\ref{sec:phaseIItemp}). The model explained substantial variance ($R^2 = 0.79$), and several categories emerged as significant predictors of return (Table~\ref{tab:regression-chatgpt}).
 
\textbf{Positive predictors.} The strongest positive predictor was \textbf{Apologies} issued by ChatGPT ($\beta = 30.75$, $p < 0.005$): turns in which the system acknowledged error and corrected course (e.g., \emph{``I apologize for the oversight in my previous response. You are correct. The actual mean of the provided scores is approximately 12.167, not 11''}). Goal-oriented academic work also predicted return, namely \textbf{Theory Application} during information seeking ($\beta = 17.52$, $p = 0.003$) and \textbf{Editing or Improving Writing} ($\beta = 9.15$, $p < 0.005$), both higher-order tasks that position ChatGPT as an academic partner (e.g., \emph{``what is the probability that at least 1 error is made''}; \emph{``make this shorter: assessment evaluates both business results and how well key business processes are managed''}).
 
\textbf{Negative predictors.} Markers of interactional strain predicted attrition. \textbf{Interaction Repair or Prompt Revision}, where students rephrased or corrected inputs after a miscommunication, had a significant negative effect ($\beta = -12.62$, $p < 0.005$), as did ChatGPT's \textbf{Stated Limitations} ($\beta = -12.03$, $p < 0.005$); both index breakdowns that pushed the cost of resolution onto the user (e.g., \emph{``no i want the full implementation based on the instructions for each method''}; \emph{``I mean difference in terms of heap/stack''}). \textbf{Frustration or Dissatisfaction} ($\beta = -11.93$, $p < 0.005$), \textbf{Follow-Up or Clarification Requests} ($\beta = -10.00$, $p = 0.002$), and \textbf{Unintended or Inconsistent Answers} ($\beta = -9.37$, $p < 0.005$) were likewise negative, indicating that disconfirming experiences, and the effort of repairing them, eroded continued engagement (e.g., \emph{``thats still wrong''}; \emph{``no but this is not working. THis IS NOT rght''}).
 
Productive academic tasks and system-led self-correction sustained use. Unresolved breakdowns, affective discontent, and perceived inadequacy predicted attrition. These patterns align with longstanding observations that affective states in learning are not uniformly destructive. \citet{baker_better_2010} show that frustration is less persistent and less learning-damaging than boredom, and \citet{kapoor_automatic_2007} demonstrate that pre-quit frustration is predictable from multimodal cues. That prediction is a useful design lever, because the most damaging negative predictors in our model are the cues of accumulating, unaddressed frustration.
 
These coefficients index a turn-by-turn repair dynamic, and it is here that RQ3 comes into focus. When a breakdown was repaired, the apology did the work the model rewards: a student counting subsets flagged an overlooked overlap (\emph{``does the 6 in both s and t not affect the ansewr''}), ChatGPT conceded (\emph{``You're right. The element 6 is in both sets''}), and when pressed on the corrected result (\emph{``why is the answer 11 choose 4 - 6 choose 4''}) it recomputed under an explicit apology. When repair failed, the turn types that load negatively here accumulated without resolution: a student debugging an assembly \texttt{phase\_2} loop cycled through \emph{``all these dont work''}, \emph{``no but this is not working. THis IS NOT rght''}, and \emph{``doesnt work, its a differnt sequence''} while ChatGPT apologized repeatedly but produced no working fix. The predictor signs thus track whether the repair loop closed, not whether errors occurred.
 
\begin{table*}[ht]
\centering
\caption{Top regression features predicting continued student use of ChatGPT. Section references in parentheses link each subcategory to its codebook entry in Appendix~\ref{sec:appendixA}.}
\label{tab:regression-chatgpt}
\begin{tabular}{p{8cm}p{6cm}rr}
\toprule
\textbf{Subcategory} & \textbf{Main Category} & \textbf{$\beta$} & \textbf{p} \\
\midrule
Apologies (\S\ref{ssec:codebook-apologies}) & ChatGPT Response Behavior & 30.75 & $<0.005$ \\
Theory Application (\S\ref{ssec:codebook-theory-application}) & Information Seeking & 17.52 & 0.003 \\
Editing or Improving Writing (\S\ref{ssec:codebook-editing-or-improving}) & Content Generation & 9.15 & $<0.005$ \\
Interaction Repair or Prompt Revision (\S\ref{ssec:interaction-repair-prompt-revision}) & Student--ChatGPT Interaction & $-12.62$ & $<0.005$ \\
Stated Limitations (\S\ref{ssec:codebook-stated-limitations}) & ChatGPT Response Behavior & $-12.03$ & $<0.005$ \\
Frustration or Dissatisfaction (\S\ref{ssec:sub_frustrate}) & Student--ChatGPT Interaction & $-11.93$ & $<0.005$ \\
Follow-Up or Clarification Requests (\S\ref{ssec:sub_followup}) & Student--ChatGPT Interaction & $-10.00$ & 0.002 \\
Unintended or Inconsistent Answers (\S\ref{ssec:uninteded-or-inconsistent-answers}) & ChatGPT Response Behavior & $-9.37$ & $<0.005$ \\
\bottomrule
\end{tabular}
\end{table*}
 
\subsubsection{Who Continues to Use ChatGPT (RQ2b)}
\label{sec:findings_survival}
 
In the Cox Proportional Hazards model, the hazard ratio (HR) represents the relative risk of the event, here disengagement from ChatGPT, associated with a given behavior. HR~$<1$ indicates reduced risk: that is, the behavior is associated with a longer time to disengagement, or greater sustained engagement. Significant categories are presented in Table~\ref{tab:cox-top-hr}, and fall into two patterns.
 
First, system breakdowns did not drive students away. \textbf{Unintended or Inconsistent Answers} carried the lowest hazard ratio (HR~$=$~0.01, $p < 0.005$). Rather than abandoning failed exchanges, students demonstrated interactional resilience, seeking clarifications, corrections, and alternative approaches, and recalibrating trust through systematic verification. One student persisted through 141 turns of ChatGPT errors by methodically testing syntax variations (\texttt{dataFrame -<- read.csv} vs.~\texttt{dataFrame <- read.csv}) to probe the system's reliability. Even ChatGPT's \textbf{Stated Limitations} (HR~$=$~0.03, $p = 0.03$) predicted retention when students adapted their prompts rather than quitting.
 
This resilience answers the resolution half of RQ3. Beyond supplying missing constraints, students resolved uncertainty by verification rather than trust: the 141-turn \texttt{read.csv} case above submitted a malformed assignment operator, a missing parenthesis, and a mismatched quote in turn, accepting only the construction ChatGPT validated across every variant. Attrition, in contrast, tracked failed repair rather than error frequency; when apologies recurred without a correct fix, the exchange stalled and the student did not return.
 
\begin{table*}[h]
\centering
\caption{Top significant ChatGPT usage categories predicting longer engagement, based on hazard ratios from the Cox PH model. HR~$<$~1 indicates reduced risk of disengagement. Section references in parentheses link each subcategory to its codebook entry in Appendix~\ref{sec:appendixA}.}
\label{tab:cox-top-hr}
\begin{tabular}{p{8cm}p{6cm}rr}
\toprule
\textbf{Subcategory} & \textbf{Main Category} & \textbf{HR} & \textbf{p} \\
\midrule
Unintended or Inconsistent Answers (\S\ref{ssec:uninteded-or-inconsistent-answers}) & ChatGPT Response Behavior & 0.01 & $<0.005$ \\
Job Application Content (\S\ref{ssec:sub_job_app}) & Content Generation & 0.01 & $<0.005$ \\
Multiple Choice or Fill-in-the-Blank Questions (\S\ref{ssec:multiple-choice-or-fill-in-the-blank-questions}) & Content Generation & 0.01 & 0.01 \\
Code Writing (\S\ref{ssec:codebook-code-writing}) & Content Generation & 0.02 & $<0.005$ \\
Theory Application (\S\ref{ssec:codebook-theory-application}) & Information Seeking & 0.02 & $<0.005$ \\
Email or Letter Generation (\S\ref{ssec:sub_email}) & Content Generation & 0.03 & 0.03 \\
Stated Limitations (\S\ref{ssec:codebook-stated-limitations}) & ChatGPT Response Behavior & 0.03 & 0.03 \\
\bottomrule
\end{tabular}
\end{table*}
 
Second, goal-oriented and iterative work predicted sustained use. \textbf{Job Application Content}, resume bullet points, cover letters, and interview responses, showed a very low hazard ratio (HR~$=$~0.01, $p < 0.005$), reflecting how goal-oriented tasks foster repeat engagement (e.g., \emph{``should i mention my social media?''}; \emph{``should i put my high school marching band on there even though im in college and in a different marching band now?''}), as did \textbf{Multiple Choice or Fill-in-the-Blank Questions} (HR~$=$~0.01, $p = 0.01$). \textbf{Code Writing} strongly predicted continued use (HR~$=$~0.02, $p < 0.005$): students prompted the assistant to generate Python or JavaScript from natural-language descriptions, frequently returning with bugs and clarification requests. \textbf{Theory Application} (HR~$=$~0.02, $p < 0.005$) was similarly engagement-sustaining (e.g., \emph{``why does a llrb tree have a max height of $2\log n$''}), and \textbf{Email or Letter Generation} (HR~$=$~0.03, $p = 0.03$) rounded out the engagement-sustaining set.

Not all interactions shape engagement trajectories equally. Behaviors tied to correction, real-world relevance, and iterative work predicted longer engagement, especially when bound to academic or professional goals. Contrary to the assumption that ChatGPT's errors discourage continued use, conversational breakdowns could deepen involvement so long as students remained motivated to repair the dialogue. The relational angle echoes \citet{skjuve_my_2021}'s longitudinal account of human--chatbot relationships, in which engagement persists through the limitations of the conversational partner so long as users can find affective exploration of the assistant.


\subsection{An Expectancy-Violation Map of Continued and Discontinued Use (RQ3)}
\label{sec:findings_evt}

The regression and survival models indicate which behaviors precede return and which lead
to attrition. They do not say why. Two negatively signed regression coefficients,
\textbf{Unintended or Inconsistent Answers} and \textbf{Stated Limitations}, sit alongside
hazard ratios far below one, so the same behavior depresses next-week volume while
lengthening time to disengagement. Expectancy Violations Theory offers a reading under which
the divergence is coherent, by separating two things our coefficients conflate, whether a
behavior fell within what the student anticipated and whether the student welcomed it.

Following \citet{burgoon_application_2016}, we treat that separation as two axes.
\emph{Expectancy} asks whether the turn fell inside the range of behavior the student had
reason to anticipate. \emph{Valence} asks whether the student appraised it favorably.
Crossing the axes yields positive violation, positive confirmation, negative confirmation,
and negative violation. Placement on the two axes is interpretive rather than modeled,
arrived at in the Tier III pass over transcripts from the most heavily engaged students
(\S\ref{sec:tier3}). We assign each subcategory from its codebook definition
(Appendix~\ref{sec:appendixA}) together with the interactions that follow it, which we read
as the observable trace of appraisal. Transcripts for the exchanges quoted below appear at
greater length in Appendix~\ref{sec:appendixC}.

Table~\ref{tab:evt-matrix} presents the map, which carries a third dimension alongside
expectancy and valence. Because what separates the four conditions behaviorally is repair,
we record in each cell the status of \emph{repair gratification}, the satisfaction of having
the exchange itself restored. We register it as \emph{delivered} where the system concedes,
\emph{not required} where nothing has broken, \emph{latent} where the student can contest an
error and the system might yet concede, and \emph{foreclosed} where the system declines
without registering the refusal as a failure.

\begin{table*}[t]
\centering\footnotesize
\caption{Expectancy-violation map of the significant predictors. Expectancy is whether the
system's turn fell inside the student's anticipated range, and valence whether the student's
next turn appraised it favorably. $\beta$ values come from the time-lagged regression
(Table~\ref{tab:regression-chatgpt}) and HR values from the Cox model
(Table~\ref{tab:cox-top-hr}). The repair-gratification column uses the same vocabulary
as Table~\ref{tab:grat-cox}.}
\label{tab:evt-matrix}
\begin{tabular}{p{2.0cm}p{3.0cm}p{2.6cm}p{2.2cm}p{2.8cm}p{2.4cm}}
\toprule
\textbf{EVT condition} & \textbf{What it looks like in the logs} & \textbf{Gratification status} & \textbf{Repair gratification} & \textbf{Model evidence} & \textbf{Engagement outcome} \\
\midrule
\textbf{Positive violation} \newline (unexpected, favorable)
& ChatGPT names its own error and supplies the correction unprompted; non-judgmental uptake of an emotional aside
& Utilitarian (the correction, and the error localized) and social (the system takes responsibility), neither of them sought
& \emph{Delivered.} The residual once the instrumental and social layers are accounted for: relational, meta-communicative, not reducible to either
& Apologies $\beta = 30.75$, $p < 0.005$; $1.8\times$ the next-largest positive coefficient, which is the purest utilitarian case (Theory Application $\beta = 17.52$)
& Strongest positive predictor of return. Closes the loop the negative-violation row opens \\[2pt]
\addlinespace
\textbf{Positive confirmation} \newline (expected, favorable)
& A usable draft, a compiling function, a correct worked answer delivered on the first pass
& Utilitarian gratification fulfilled and externally verifiable. Social gratification in self-presentation tasks
& \emph{Not required.} Nothing has broken, so none is at stake
& Theory Application $\beta = 17.52$, HR~$=$~0.02; Editing/Improving Writing $\beta = 9.15$; Code Writing HR~$=$~0.02; Job Application Content HR~$=$~0.01; MCQ HR~$=$~0.01; Email/Letter HR~$=$~0.03
& Durable, but a smaller per-turn effect than the positive violation \\[2pt]
\addlinespace
\textbf{Negative violation} \newline (unexpected, unfavorable)
& Confidently wrong arithmetic, contradictory answers across turns, code that does not run
& Utilitarian gratification disrupted at the point of delivery
& \emph{Latent.} The student's contest opens the loop; available the moment the system concedes
& Unintended/Inconsistent Answers $\beta = -9.37$, HR~$=$~0.01; Frustration $\beta = -11.93$; Prompt Revision $\beta = -12.62$; Follow-Up $\beta = -10.00$
& Depresses next-week volume, yet retains students whose repair loop closes into the positive-violation row above \\[2pt]
\addlinespace
\textbf{Negative confirmation} \newline (expected, unfavorable)
& A refusal the student has learned to anticipate, such as no browsing, no real-time data, no personal experience
& Gratification withheld before any attempt. The workaround cost falls entirely on the student
& \emph{Foreclosed.} A refusal is not registered as a failure, so none is offered
& Stated Limitations $\beta = -12.03$, HR~$=$~0.03
& Damages engagement about as much as the negative violation does, with no repair pathway available \\
\bottomrule
\end{tabular}
\end{table*}

\subsubsection{Positive violations: the system makes the correction}
ChatGPT gives a wrong or unanticipated answer, the student contests it, and ChatGPT
concedes and corrects. A student working through an automata theory exercise asked how many
of four statements about nondeterministic and deterministic finite automata were true.
ChatGPT marked three of the four true, then reported the total as two. The student caught
the mismatch and pushed back, asking \textit{``isn't that 3 statements you said were
true''}. ChatGPT replied \textit{``You are correct. I apologize for the error. The number of
true statements is 3''}, and the student wrote \textit{``okay thank you''} and moved on.
Appendix~\ref{ssec:appC_repair_closes} sets out two further instances at greater length.

The same shape appears outside instrumental work. A student who had disclosed distress
about seeking therapy wrote \textit{``i just feel sort of pitiful and weak looking for overt
help like that. i don't like the idea of having to give up my time every week just to make
myself functional''}. ChatGPT neither redirected them to a helpline nor pressed the case for
therapy. It contested the premise, replying that \textit{``seeking help when you need it is
actually a sign of strength''}, widened the frame by noting that \textit{``therapy isn't just
for people who are struggling to function''}, and then returned the decision to the student,
adding that \textit{``if you're not ready to see a therapist right now, that's okay''}. The
student answered \textit{``thank you''}.

In both cases the system did something a chatbot is not assumed to do well, and the student's next turn registers a favorable appraisal. This is where the apology coefficient
($\beta = 30.75$) lives, and its size relative to every task-completion predictor is what EVT anticipates when a favorable act arrives from outside the expected range. The apology does
three things at once. It carries the correction and localizes the error, which is utilitarian. It takes responsibility for the failure, which is social. It restores the exchange itself. We name that third layer \emph{repair gratification} and explore it in \S\ref{sec:disc_repair}. In this cell repair gratification is \emph{delivered}.

\subsubsection{Positive confirmations: expected gratification delivered}
The engagement-sustaining tasks in the Cox model are positive confirmations. Code compiled (\textbf{Code Writing}), students drew usable bullet points from cover letter requests (\textbf{Job Application Content}), and \textbf{Multiple Choice or Fill-in-the-Blank Questions} resolved. One student closed a long debugging exchange by checking the shape of the
solution rather than its content, asking \textit{``or I can save this script to a file and
reference this file in index.html, correct?''} before writing \textit{``thank you for your
help.''} Another treated ChatGPT as a self-presentation advisor, asking \textit{``should i
mention my social media?''} (Appendix~\ref{ssec:appC_sustaining}). These prompts recruit
utilitarian gratification, and in the job-application case a social gratification of rehearsed
self-presentation as well.

What distinguishes them is verifiability. The student can tell, from outside the
conversation, whether the gratification was delivered. Their hazard ratios are low and their
regression coefficients positive, but none approaches the apology. While this confirmation might retain users, 
violation moves the needle on (non)use. Repair gratification is \emph{not required} here, because nothing
has broken.

\subsubsection{Negative violations: arousal that cuts both ways}
\textbf{Unintended or Inconsistent Answers} carry the split signature that motivated this
section, $\beta = -9.37$ against HR~$=$~0.01. A negative violation is arousing
\citep{burgoon1978communication}. It pulls the student's attention onto the system and away
from the task \citep{burgoon1988nonverbal}, which is why the following week's volume drops.
It also makes the system newly salient as an object of scrutiny, which is why the student
stays. Verification is the visible form of that scrutiny \citep{acosta2024analysis}.

A single 141-turn session shows why this cell is not simply a record of system error
(Appendix~\ref{ssec:appC_negviol}). At turn 119 ChatGPT rejected the valid identifier
\texttt{.dataFrame} on the ground that R variable names cannot begin with a dot and must begin
with a letter or an underscore. Both halves of that rule are inverted, since R permits a
leading period and forbids a leading underscore. The student saw this and said so, in the
words \emph{``its fine its right''}, and ChatGPT read the correction as praise, thanked the
student for it, and offered further help. The act was unexpected and unfavorable, and the
loop never closed, not because the correction was refused but because it was never registered
as a correction. This is what a negative violation looks like when it cuts against the user.
The student leaves holding a true belief that the system has told them is false.

The same session supplies the positive case, which is legible only against the negative one.
Across turns 15 to 30 the student pressed an incorrect answer on a null-hypothesis item and
ChatGPT reversed itself three times under that pressure, ending on the student's wrong choice
(Appendix~\ref{ssec:appC_capitulation}). By turn 120, then, the student had evidence gathered
in this very conversation that assertion moves the system. Roughly a hundred turns later they
ran the identical play on a malformed file extension, asserting \emph{``ur wrong''} and
resubmitting the same line three times, and the verdict did not move
(Appendix~\ref{ssec:appC_resistance}). Holding a correct answer against a user who has already
succeeded in moving one is both unexpected and favorably evaluated, which places it in the
positive-violation cell and locates the gratification in the system's steadiness rather than
in its agreeableness.

Negative violations convert into positive violations only if the system closes the loop.
When it did not, the turns accumulated. One student, twelve turns into debugging a single R function, wrote \textit{``That does not work at all. Adding +13 makes the problem significantly worse''} and later \textit{``This still does not work at all. It does not produce 60605''}. Another, debugging lines of assembly code, cycled through \textit{``all these dont work''} and \textit{``doesnt work, its a differnt
sequence''} while ChatGPT apologized on every turn and produced no working fix
(Appendix~\ref{ssec:appC_repair_fails}). Prompt revision, follow-up clarification, and
frustration are the student's side of an open loop. They are negative predictors not because
repair is costly but because unfinished repair is. Repair gratification is \emph{latent}
throughout this cell. It becomes available the moment the student contests the error and is
realized only if the system acknowledges the error and provides a correction, which is what moves an exchange into the
positive-violation cell above.

\subsubsection{Negative confirmations: the refusal students saw coming}
\textbf{Stated Limitations} are the one system behavior students could easily
recognize. They account for 32\% of \emph{ChatGPT Response Behavior}, itself 22\% of
annotated turns (Appendix~\ref{ssec:codebook-chatgpt}), so within a few encounters a student knows
both when a refusal is likely and roughly how it will be worded.

Anticipation leaves two options, and the split between them produces this cell's mixed
signature. Some students absorbed the cost and rewrote the request. One asked ChatGPT to
answer an interview question in its own voice, \textit{``answer: `why are you qualified for the
position of community assistant for your resident building?'\,''}, and was told \textit{``As an
AI language model, I do not have personal qualifications, experiences or emotions''}. The
student reframed the same question as a request for advice, \textit{``how can i answer \dots''},
received a usable outline, then asked for a sample response and got one. The refusal cost two
turns and nothing else. Reformulation of this kind keeps students in the conversation and is
what holds the hazard ratio at 0.03. However, other students did not reformulate. After ChatGPT declined
to compare two pages of a coloring book because it could not see the images, one student's
entire reply was \textit{``dumbass''}, and the conversation ended there after two turns.
Departures of that kind are what might explain the reduced future use with a regression coefficient of $-12.03$.

The comparison across the two negative cells is the analytically interesting one. If
surprise were having more of a negative effect for future use, \textbf{Unintended or Inconsistent Answers} should hurt more. \textbf{Stated Limitations} carries the larger absolute regression coefficient
($-12.03$ against $-9.37$) and a hazard ratio of the same order (0.03 against 0.01). Being
failed in a way one saw coming is no better than being failed by surprise.
\citet{burgoon_application_2016} report a comparable null in a laboratory study of embodied agents, where negative violations proved no more damaging than negative confirmations. Our two negative cells point in the same direction. This suggests that their findings are not limited to embodied agents, but seem to apply to text-only AI assistants.


Our findings suggest that well-defined \textbf{Stated Limitations} do not represent a safe interaction with the user that will increase use. What separates the two negative cells from the positive-violation cell is not whether the student anticipated the failure, but rather how the system responded. \textbf{Unintended or Inconsistent Answers} become errors from the system's side as soon as the student contests them, and that is what makes an \textbf{Apology} available. \textbf{Stated Limitations} on the other hand suggest that the system has behaved as designed, so nothing triggers repair and the student is left to supply the workaround unaided. Repair
gratification is therefore not merely absent from the negative-confirmation cell but structurally foreclosed by it, which is why a well-formed refusal can cost as much engagement
as a wrong answer.

\section{Discussion}
\label{sec:discussion}
 
This study gives an empirical, longitudinal view of how undergraduates integrate ChatGPT into self-directed learning. We organize the discussion around three themes: what students do and who carries the repair work when interactions break down (\S\ref{sec:disc_rq1}); \emph{repair gratification}, the mechanism by which system-issued apologies become the strongest predictor of continued use (\S\ref{sec:disc_repair}); and \emph{temporal incorporation}, the uneven process by which different uses become, or fail to become, embedded in academic rhythms (\S\ref{sec:disc_temporal}). EVT, within the human--machine communication agenda \citep{guzman_artificial_2020}, anchors the interpretation alongside SDL and UGT.
 
\subsection{A Sociotechnical Learning Ecology and Co-Regulated Repair}
\label{sec:disc_rq1}
 
Our taxonomy shows students using ChatGPT as a ``Swiss-Army-knife'' of academic support,
shifting within a single session between explainer, writing coach, job-search assistant, and
emotional sounding board, from academic tasks \citep{kim2024chatgpt} to personal decisions
\citep{von2023artificial}. This breadth advances work confined to single domains such as
computer-science education or EFL writing \citep{uddin2024chatgpt, kim2024designing,
han2024recipe4u} and reflects the convergence of utilitarian, hedonic, and social
gratifications that UGT (\S\ref{sec:rw_ug}) predicts of active users \citep{mclean2019hey}.
\citet{xu_tool_2024} find that functional and relational use reinforce one another over time,
and students who folded socio-emotional turns (\textit{``yo chatgpt i am wildly unhappy''})
into instrumental sessions were doing what predicts sustained functional engagement in voice
assistants. \citet{skjuve_my_2021} show such relational dynamics deepening even in text-only
systems, and we extend that from a companion chatbot to an academically positioned one.

\citet{haugeland_understanding_2022} distinguish task-led conversations, which pursue a
defined outcome, from topic-led ones, which range more freely and better prompt reflection.
Repair sits across that distinction rather than on one side of it. \textbf{Unintended or
Inconsistent Answers} can only be recognized as such where there is a defined outcome to fall
short of, so repair is triggered in task-led work. What it delivers is the relational and
reflective good \citet{haugeland_understanding_2022} associate with topic-led work, because
contesting the answer makes the exchange itself the temporary topic. Repair is therefore the
mechanism by which a task-led conversation becomes briefly topic-led and then returns, which
is why \textbf{Apologies} outperform every task-completion predictor in our models
($\beta = 30.75$ against $\beta = 17.52$ for \textbf{Theory Application}). They pay in both currencies
at once. \textbf{Stated Limitations} pay in neither ($\beta = -12.03$). They withhold the
outcome, so there is no task-led return, and because the system has behaved as designed there
is nothing for the student to contest, so the exchange never becomes the topic
(\S\ref{sec:design}).
 
Our findings show students managing ambiguity of two kinds, cognitive confusion over content
and epistemic mistrust in responses \citep{rohmann2023case}. Rather than abandon the tool,
they met it with clarification requests, follow-ups, and revised prompts \citep{liu2021ai,
aquilino2024trust, braggaar2023breakdowns, Ashktorab_2019}. This complicates the assumption
that breakdowns inevitably cause disengagement \citep{zeng2024chatbot,
folstad2021investigating}. Ambiguity is often read as a barrier to trust \citep{lee2024toward}
and to long-term use \citep{babel2022will}, but ambiguity that students had the skill or
motivation to resolve could instead deepen involvement, echoing findings that reciprocity
\citep{lee_enhancing_2017} and equity \citep{gockley2005designing} sustain use.

Recovery was not evenly distributed. \citet{rapp_how_2025} attribute such variation to
expertise, finding that experienced users absorb an unpredictable response into a
machine-frame and continue while less experienced ones do not. Our logs cannot test that
account, since they carry no measure of prior experience. What they do show is that students
differed in how far they carried a broken exchange, and that the task genres associated with
the longest engagement were technical ones, with \textbf{Code Writing }and \textbf{Theory Application} among the
lowest hazard ratios (Table~\ref{tab:cox-top-hr}). Whether that reflects who undertakes such
work, the demands of the tasks themselves, or the unambiguous failure signal a script provides,
logs alone cannot say.

These practices reframe self-directed learning as \emph{co-regulated}. Students set explicit
goals (\textit{``give me a practice s2 cognition test''}), checked their reasoning
(\textit{``isnt it B? Not sure''} and \textit{``am i correct in saying that they are the
same?''}), and reframed prompts after a misunderstanding, extending Song and Hill's
individual-trait model \citep{song2007conceptual} toward a dialogic and emergent process
\citep{li2024reconceptualizing}. Regulation ran in both directions, though not always successfully. ChatGPT acknowledged its
own errors and opened routes to correction, and where it followed through the loop closed in
a turn or two. Where it did not, the acknowledgment recurred without a working fix and the
exchange stalled (Appendix~\ref{ssec:appC_repair_fails}). In both cases it is acting as a
socio-technical partner rather than a static tool, which extends to open-ended LLM use a
logic long present in adaptive tutors \citep{dmello_gaze_2012, kapoor_automatic_2007}. The repair
loops that did close leave a gratification unaccounted for. A student met with an acknowledgment and a correction gains more than the corrected answer,
and none of the four gratifications set out in \S\ref{sec:rw_ug} accounts for it.

\subsection{Repair Gratification: Accountability over Accuracy}
\label{sec:disc_repair}

Continued engagement depended on gratification \emph{fulfillment} rather than \emph{seeking},
and reliability varied by task. The strongest positive predictors were interactions in which
utilitarian gratifications were reliably and \emph{verifiably} satisfied.
\textbf{Theory Application} ($\beta = 17.52$), \textbf{Job Application Content}
(HR~$=$~0.01), and \textbf{Code Writing} (HR~$=$~0.02) all supply an external check, since
code compiles or fails, cover letters draw feedback, and explanations clarify or do not. The
negative predictors index gratification \emph{disruption}. \textbf{Interaction Repair or
Prompt Revision} ($\beta = -12.62$), \textbf{Frustration or Dissatisfaction}
($\beta = -11.93$), and \textbf{Unintended or Inconsistent Answers} ($\beta = -9.37$) mark
points where interaction cost exceeded benefit, and where ChatGPT is better cast as a
socio-technical actor co-constructing learning than as a cognitive tool
\citep{schulz2025generative}.

Table~\ref{tab:grat-cox} intersects the engagement-sustaining predictors of
Tables~\ref{tab:regression-chatgpt} and~\ref{tab:cox-top-hr} with the gratification each
recruits, the EVT condition under which it is fulfilled, the status of repair
gratification in each case, and how it broke down in our logs. Two regularities follow.
Every subcategory with a hazard ratio at or below 0.02 gives the student an external check,
whether a compiler, an answer key, a rubric, or a derivation they can redo, and verifiability
is what converts a delivered answer into a fulfilled gratification. Repair gratification
then varies in degree across the whole table rather than sitting only in the
\textbf{Apologies} row. It is \emph{delivered} where the system concedes, \emph{latent}
wherever the student can contest an error and the system might yet concede, and
\emph{foreclosed} wherever the system announces a limitation it does not register as a
failure. The two rows in which it is foreclosed are also the two in which the workaround cost
transfers wholly to the student. Neither regularity runs through task success, which is why
a gratifications account has to be extended past the task.

\begin{table*}[t]
\centering\footnotesize
\caption{Engagement-sustaining predictors intersected with the gratification each recruits, the EVT condition under which it is fulfilled, the status of repair gratification in the same vocabulary as Table~\ref{tab:evt-matrix}, and the point at which it failed. Evidence combines the Cox model (Table~\ref{tab:cox-top-hr}, where HR below one indicates reduced risk of disengagement) and the time-lagged regression (Table~\ref{tab:regression-chatgpt}, where $\beta$ is the change in following-week volume). Apologies does not appear among the significant Cox predictors and is entered on its regression coefficient, listed first as the strongest positive predictor in either model.}
\label{tab:grat-cox}
\begin{tabular}{p{2.7cm}p{2.4cm}p{1.9cm}p{2.5cm}p{2.9cm}p{2.5cm}}
\toprule
\textbf{Subcategory (model evidence)} & \textbf{Gratification status} & \textbf{EVT condition when fulfilled} & \textbf{Repair gratification} & \textbf{Where it broke down} & \textbf{Consequence for incorporation} \\
\midrule
Apologies ($\beta = 30.75$) & Layered. Utilitarian (correction supplied, error localized), social (the system takes responsibility), and \emph{repair gratification} as the residual. None of the three sought & Positive violation & \emph{Delivered.} System-initiated, since this \emph{is} the repair move, and the loop closes when a correction rides with it & Acknowledgment issued without a working correction, so the apology ritualizes (\textit{``no but this is not working. THis IS NOT rght''}) & Strongest positive predictor of return. Where only the social layer lands, the apology ritualizes and predicts termination \\
\addlinespace
Unintended or Inconsistent Answers (HR~$=$~0.01, $\beta = -9.37$) & Utilitarian, sought by the student and disrupted in the system's turn & Negative violation, convertible to positive violation & \emph{Latent.} The student contests and opens the loop, which converts only if the system concedes and corrects & Apologies recurring without a working fix (\textit{``all these dont work''}) & Retention where the repair loop closed, mid-semester termination where it did not \\
\addlinespace
Job Application Content (HR~$=$~0.01) & Utilitarian and social (rehearsed self-presentation) & Positive confirmation & \emph{Foreclosed} at the stated limitation, so the student recasts the prompt unaided & Limitation stated on personal-experience grounds (\textit{``I do not have personal qualifications''}) & Survives if the student recasts the prompt as advisory \\
\addlinespace
Multiple Choice or Fill-in-the-Blank Questions (HR~$=$~0.01) & Utilitarian, with an answer key as external check & Positive confirmation & \emph{Latent}, and immediately so. The key exposes the error, the student contests, the system concedes & Confident selection of a wrong option (\textit{``thats wrong''}) & High retention, since the error surfaces immediately and can be repaired \\
\addlinespace
Code Writing (HR~$=$~0.02) & Utilitarian, with a compiler as external check & Positive confirmation & \emph{Latent} and iterative. The compiler localizes the fault and repair cycles until it runs & Code that runs but produces the wrong output (\textit{``it does not produce 60605''}) & Retention while the student can localize the fault, attrition when no fix arrives \\
\addlinespace
Theory Application (HR~$=$~0.02, $\beta = 17.52$) & Utilitarian and epistemic, being understanding rather than output & Positive confirmation & \emph{Latent}, student-side. The student re-derives, so repair \emph{is} the learning activity & Plausible but wrong derivation, checkable only by redoing the work & Retention, because the checking \emph{is} the learning activity \\
\addlinespace
Email or Letter Generation (HR~$=$~0.03) & Utilitarian and social (managing a relationship) & Positive confirmation & \emph{Latent} but rarely invoked. Register mismatch is repaired by the student's own rewrite & Register or tone mismatch requiring rewriting & Durable but narrow, confined to writing-adjacent moments \\
\addlinespace
Stated Limitations (HR~$=$~0.03, $\beta = -12.03$) & Utilitarian sought and none obtained. Withheld in the system's turn before any attempt & Negative confirmation & \emph{Foreclosed.} The system does not register a stated limitation as a failure, so none is offered & Workaround cost transferred wholly to the student & Retention only where the student can supply the missing context \\
\bottomrule
\end{tabular}
\end{table*}

The strongest result complicates that binary. \textbf{Apologies} were the single strongest
positive predictor of continued use ($\beta = 30.75$, $p < 0.005$), outweighing every
task-completion predictor. We term this \emph{repair gratification}, the satisfaction of
having the exchange itself restored, over and above the correction the apology carries and
the responsibility it takes. Acknowledgment of failure re-established alignment between
behavior and expectation, so apologies worked not as politeness but as functional repair
restoring conversational continuity and psychological safety \citep{brandtzaeg2018chatbots,
luger2016thegulf}. Responsiveness rather than infallibility sustained commitment, and
students who persisted through inconsistent responses showed lower attrition.

An apology is not a single gratification but a layered one, and separating the layers is
what licenses the construct. Instrumentally, the acknowledgment carries or precedes a
correction and localizes the error, telling the student which turn to distrust, and that
layer is utilitarian. Relationally, a system that takes responsibility for its own failure
performs the reciprocity \citet{lee_enhancing_2017} find predicts continued use more strongly
than self-disclosure, and that layer is social. Neither accounts for the size of the
coefficient. A correct answer delivered on schedule, the purest utilitarian case in our
models, earns barely more than half the apology's weight (\textbf{Theory Application}
$\beta = 17.52$ against $30.75$), and acknowledgment offered without a correction does not
retain at all. \emph{Repair gratification} names the increment that remains once the
instrumental and social layers are accounted for, the meta-communicative satisfaction of
having the interaction itself restored rather than merely the answer corrected.

Three strands anchor the mechanism, all within HMC's treatment of AI as a communicative
partner \citep{guzman_artificial_2020}. An apology is an economical reciprocity, returning
acknowledgment of the student's correction at low cost. It is also a positive violation in
the sense mapped in \S\ref{sec:findings_evt}, since students do not expect a system to admit
fault, and \citet{burgoon_application_2016} show that such violations move relational
judgments further than meeting expectations does. And \citet{pan_humanai_2025} find that
more agentic AI, with an autonomous source and first-person language, reads as more
trustworthy and likable, so a system-issued apology pulls all three levers at once. 

However, not every apology repairs, and our logs separate three grades.
\emph{Acknowledgment with correction} is the full move, in which the system concedes and
supplies the fix in the same turn, as in the finite-automata exchange quoted in
\S\ref{sec:findings_evt}. \emph{Acknowledgment alone} concedes the error without resolving
it, leaving the loop open and the repair cost with the student. \emph{Ritualized
acknowledgment} is the degenerate case, in which the apology recurs as a conversational
reflex. In one ninety-one-turn debugging conversation ChatGPT apologized in fifteen separate
turns without ever producing a working fix, and the student abandoned the exchange
(Appendix~\ref{ssec:appC_repair_fails}). The coefficient rewards the first grade, but the
third is indistinguishable from it at the level of a code, which is why apology frequency is
a poor design target on its own. What retains students is acknowledgment that carries a correction. Because the covariate is a count of apologies rather than of repaired exchanges, the coefficient pools all three grades and estimates their average association with return. The ritualized case is therefore not an exception the model absorbs but a distinction it cannot draw, and separating the grades would require coding whether a correction accompanied the acknowledgment. We therefore read the coefficient as evidence that repair sequences occur and matter, not as an estimate of the return on any particular number of apologies.

A fourth case falls outside these grades because no apology is issued at all. In
\S\ref{sec:findings_evt} a student corrected a wrong claim about R naming rules and the
system read the correction as praise, so the loop never opened. Repair depends on uptake
before it depends on accuracy, since a correction the system does not register as one cannot
be conceded. The same session showed the converse, where the system held a correct answer
against three assertions that it was wrong, and the gratification lay in its steadiness
rather than in its agreeableness. Neither case would appear in a count of
\textbf{Apologies}, which is a further reason to treat apology frequency in and of itself as a weak design target.

Repair gratification extends UGT in two ways. First, it adds a \emph{meta-communicative} gratification
absent from media-gratification frameworks, derived from interaction quality rather than
instrumental output, and consisting in being heard and appropriately answered when a
misunderstanding occurs. Second, it also recasts fulfillment and disruption as ongoing \emph{repair
cycles} rather than binary states, since a disruption later repaired can strengthen
engagement more than uninterrupted fulfillment, because repair demonstrates the relationship's
resilience. This is the EVT logic of Table~\ref{tab:evt-matrix} in gratification terms, with
the positive-violation cell dominating the positive-confirmation cell, and it extends UGT
from adoption to \emph{retention}. Students return when gratification circuits close, either
by initial fulfillment or by post-disruption repair, and they leave when disruptions go
unrepaired.

\subsection{Temporal Incorporation into Academic Rhythms}
\label{sec:disc_temporal}

The survival results invite a longer-horizon reading. Students pass through familiarization,
after which utility rises, not from novelty but from a growing ability to appropriate the
system to their needs \citep{long2024novelty}. Where the session-grain vocabulary of EVT runs
out, domestication research supplies a longer measure. \emph{Incorporation} is the phase at
which a technology is embedded into temporal rhythms, as students integrate ChatGPT into study
routines, homework workflows, and exam preparation
\citep[pp.~21--22]{silverstone1992information}. It is therefore legible in \emph{when} a tool
is used.

Figure~\ref{fig:timeseries} shows the incorporation signature at corpus level, with
activity surging during midterms and finals and falling away during breaks. ChatGPT was woven
into the rhythms of academic life as ``mundane data'' \citep{pink2017mundane}. Students who
continued did not all continue in the same way. Some sustained a long but thinning presence
across the observation window, others intensified within a shorter one, and our Tier III
sampling operationalizes that difference (Appendix~\ref{sec:appendixB}). Both are forms of
incorporation, and total volume distinguishes neither.

Incorporation was also uneven across use genres. \textbf{Information Seeking} and
\textbf{Content Generation} prompts that yielded verifiable, task-bound value showed the
strongest retention and lodged most deeply in workflows. \textbf{ChatGPT Response Behavior}
cut both ways. \textbf{Apologies} catalyzed incorporation, tying repair gratification to
temporal embedding, whereas \textbf{Stated Limitations} predicted disengagement unless the
student adapted. Table~\ref{tab:grat-cox} locates that adaptation, since a student who
reformulates keeps the genre alive while one who answers \textit{``dumbass''} and closes the
window does not.

Interactional strain behaved differently. \textbf{Interaction Repair or Prompt Revision}
($\beta = -12.62$), \textbf{Frustration or Dissatisfaction} ($\beta = -11.93$), and
\textbf{Follow-Up or Clarification Requests} ($\beta = -10.00$) all predicted reduced return.
They mark high in-the-moment engagement while signaling the friction of managing the system,
where perceived costs outran benefits, recalling \citet{baker_better_2010} on persistent
boredom, in which apparent engagement masks drop-off. \textbf{Stated Limitations}
($\beta = -12.03$) were negative but not uniformly damaging, since students who reformulated
or shifted goals continued while those left to supply the workaround unaided did not. This
gives incorporation an observable failure mode. Use that terminates mid-semester, rather than
tapering at a natural break and resuming afterwards, marks a tool that never settled into an
academic rhythm at all, and interactional strain is what distinguishes those trajectories.

These micro-regulations accumulate into distinct patterns that emerge through students' agency
rather than technological determinism \citep{bakardjieva2005internet}. Accounts of how mobile
devices \citep{marika_et_al_2010} and voice assistants \citep{Brause15042020} settle into
daily life extend naturally to AI in dialogue, where the work of settling in is a continuous
relational and symbolic negotiation carried out through conversation itself.

\section{Design Implications}
\label{sec:design}

Our empirical findings suggest specific interaction patterns that could inform conversational AI design and institutional policy for educational contexts. Rather than prescribing features or restrictions, we offer recommendations derived from observed student practices grounded in the use-genre diversity we documented.

\subsection{Repair Scaffolding Over Accuracy Optimization}
\label{sec:design_repair}
The strongest positive predictor of continued engagement was ChatGPT's apologies ($\beta = 30.75$) rather than successful task completion. This inverts conventional design priorities. For conversational AI, \emph{how systems fail} may matter more than \emph{whether they fail}. The finding aligns with research emphasizing responsive accountability over flawless performance \citep{brandtzaeg2018chatbots, luger2016thegulf}. It is also what \citet{burgoon_application_2016} predict, since positive expectancy violations shape perceptions of connectedness and dependability disproportionately, and since simply meeting expectations carries no such guarantee. Table~\ref{tab:evt-matrix} adds a second design lesson from the same source. Anticipated failure was no less damaging than surprising failure, so a system cannot make a refusal safe by making it predictable. We propose \textbf{graduated repair patterns} as a design concept:

\begin{itemize}
\item \textbf{Uncertainty Signaling}: Systems should distinguish between confident and hedged responses through linguistic and visual markers before errors occur, addressing gaps in AI tacit knowledge \citep{jarrahi2022artificial} rather than only acknowledging uncertainty post-failure.

\item \textbf{Repair Initiation}: On detecting user frustration through follow-up clarification requests or prompt-revision patterns, systems could proactively initiate repair (``I may have misunderstood, would it help if I approached this differently?'') rather than waiting for explicit correction. \citet{kapoor_automatic_2007} demonstrate that pre-quit frustration is predictable from multimodal cues, so the prediction infrastructure exists. What is missing is a design move that uses it to initiate repair rather than only to log affect.

\item \textbf{Failure Transparency}: Our observation that students who encountered unintended or inconsistent answers showed lower disengagement risk (HR~$=$~0.01) when they persisted suggests that visible failure with transparent recovery may build trust more effectively than invisible filtering or confident errors.
\end{itemize}

The retention dynamics we document also raise the question of what kind of sustained engagement is desirable. Recent work on AI-enabled wellbeing apps offers a parallel. \citet{orji_revitalizing_2025} design and evaluate \emph{SmileApp}, an mHealth application combining AI with persuasive technology to promote positive affect, and show that theory-informed design can shift users toward desired emotional states over a 14-day in-the-wild study. The substantive domain differs from ours, but the design grammar transfers. Engagement is a means to a learner state rather than an end in itself, and design should be explicit about which state it is optimizing for. Our repair-scaffolding and mode-aware recommendations treat sustained ChatGPT use as valuable insofar as it supports learning and well-being, not as a retention metric.

\subsection{Supporting Mode Transitions and Gratification Shifts}
\label{sec:design_modes}
Students shifted fluidly between task-focused work, emotional expression, and collaborative repair, often within a single session. This movement across distinct use genres \citep{bakardjieva2005internet} suggests that conversational AI should support \emph{mode transitions} rather than force a single interaction frame \citep{collins1988computer}. \citet{haugeland_understanding_2022} find that task-led conversations strengthen pragmatic quality whereas topic-led conversations strengthen anthropomorphism and hedonic quality, so mode design is consequential rather than ornamental. Systems should adapt as gratification needs shift from utilitarian to hedonic to social:

\begin{itemize}
\item \textbf{Contextual Mode Detection}: Systems could detect genre shifts (e.g., from task-focused to emotional expression) and adapt response patterns, offering validation when students express frustration rather than immediately redirecting to task completion.

\item \textbf{Explicit Mode Selection}: Drawing on students' observed practice of framing interactions (\textit{``yo chatgpt i am wildly unhappy''}), systems could offer explicit mode toggles (Study Mode, Vent Mode, Debug Mode) that set appropriate response expectations and scaffold meta-cognitive awareness of interaction goals \citep{self1988student}.
\end{itemize}

The diversity of use genres also carries policy implications: blanket institutional prohibitions or permissions fail to recognize that different genres serve different educational purposes \citep{wester2024ai}. Using ChatGPT to draft a professional email is qualitatively different from generating an essay for submission, yet both fall under ``ChatGPT use.''

\subsection{Supporting Epistemic Vigilance}
\label{sec:design_vigilance}
Students systematically tested ChatGPT through verification queries. That trust-calibration pattern suggests designing for \emph{productive skepticism} rather than trying to eliminate doubt. It matters because most students lack the knowledge to build complex AI mental models \citep{johnson_yu_et_al_24}, an asymmetry sharpened by \citet{rapp_how_2025}'s finding that novice and expert LLM users diverge sharply when a model misbehaves:

\begin{itemize}
\item \textbf{Verification Affordances}: Integrating explainable-AI-for-education (XAI-ED) features \citep{KHOSRAVI2022100074} could scaffold the epistemic vigilance students already practice, offering tools such as ``Would you like me to show my reasoning?'' or ``Here are sources you could check'' to support informed judgment.

\item \textbf{Confidence Calibration Feedback}: Systems could provide calibration information (``I'm more reliable on X than Y topics based on my training'') supporting development of accurate mental models. Ethical nudges about citation practices and originality \citep{lara2024socratic} could further promote critical engagement.

\item \textbf{Metacognitive Reflection Prompts}: Prompts such as ``What did you learn from this?'' or ``Would you like to evaluate another perspective?'' \citep{ponomarioviene2024solutions} can encourage self-assessment and deeper engagement, reinforcing SDL by scaffolding autonomy and reflective practice \citep{song2007conceptual, hobert2019you}.
\end{itemize}

\section{Policy Implications: Toward Participatory AI Literacy}
\label{sec:policy}

Students are already developing folk strategies and workarounds. Repair rituals, trust-calibration heuristics, and genre-switching tactics represent sophisticated practical knowledge about human--AI collaboration. These lived experiences should inform platform design and institutional policy-making rather than be dismissed as ad hoc coping mechanisms.

Digital-literacy curricula should extend beyond effective prompt crafting to include strategies for recovering from breakdowns and managing failed or ambiguous interactions \citep{baidoo2023education, li2024impact}. Because \emph{relational AI literacy} matters, curricula could offer task-focused \citep{schulz2025generative} and experiential \citep{gnoth2025supporting} learning in which students engage with scenarios they will meet when augmenting their capabilities through AI partnership \citep[p.~4]{jarrahi2022artificial}.

Co-designing features with educators, and incorporating student perspectives, can help ensure alignment with ethical, pedagogical, and contextual values \citep{medeiros2021can, fitzpatrick2017delivering}. Research demonstrates that fostering AI literacy is essential for both preservice teachers facing technological anxieties and students seeking clearer institutional guidance \citep{bilal2025guest}. Hands-on interventions, including targeted literacy training and scaffolded GenAI exposure, can increase confidence, promote responsible use, and mitigate apprehension, supporting more meaningful and equitable educational outcomes.

The multi-stakeholder ethics of generative AI in higher education deserve particular attention. \citet{han2026ethical} analyze an adjacent case of AI-based monitoring of student learning and identify three social factors that erode accountability. Students' perspectives are under-represented in adoption decisions. Developers make pragmatic compromises under organizational pressure. Adopting institutions have limited AI literacy. All three translate to undergraduate generative-AI use. The students whose logs we analyze are rarely consulted in institutional policy-making about ChatGPT. Developers' decisions about refusal patterns, apology phrasing, and knowledge cut-offs propagate into classroom experiences that few faculty understand. Institutional decision-makers often lack the AI literacy to assess the trade-offs. Our repair-gratification finding sharpens the point. If a system's apologetic behavior is the strongest predictor of student retention, then platform-level choices about how and how often to apologize are pedagogical choices, and institutions adopting such platforms should engage with them as such. Designing AI systems, and adoption processes, therefore requires attention to the moral economy of academic life as well as to technical effectiveness.

\section{Limitations and Future Work}
\label{sec:limitations}
While our study provides a rich, log-based understanding of how undergraduate students engage with ChatGPT across academic and personal contexts, it has limitations that suggest avenues for future inquiry. Most notably, our analysis relied entirely on interaction logs. Although this captures real-world use at scale, it does not include direct engagement with participants through interviews or surveys; we were therefore unable to probe students' intentions, interpretations, or emotional states beyond what was observable in their prompts and ChatGPT's responses. This limits our ability to fully contextualize user motivations, evolving expectations, and perceptions of system trustworthiness or value over time. In particular, the asymmetry that \citet{rapp_how_2025} document between expert and novice users' interpretations of LLM breakdowns is visible only indirectly in the logs. Interview methods are needed to triangulate the mental-model dynamics behind the regression patterns. The same caution applies to our EVT map. We infer expectancy and valence from the student's next turn rather than measure them, as \citet{burgoon_application_2016} do with post-task instruments. Assigning subcategories to quadrants is therefore an interpretive reading of behavioral traces, and a study pairing donated logs with expectancy and valence ratings would test it directly.

Future work should complement log analysis with qualitative methods such as in-depth interviews or diary studies, as well as quantitative instruments like surveys to triangulate behavioral findings. Co-design sessions with students could provide an opportunity to collaboratively reinterpret usage patterns, uncover latent needs, and generate user-informed design ideas. Such participatory methods can deepen our understanding of how ChatGPT and similar tools might be better integrated into educational ecosystems in ways that align with student values, workflows, and goals. Multimodal extensions of the kind \citet{cosentino_exploring_2025} pursue in embodied learning environments, and longitudinal relational designs of the kind \citet{skjuve_my_2021} pioneer for companion chatbots, would together help adjudicate whether the repair-gratification mechanism we identify is specific to text-only, academically positioned LLMs or generalizes across conversational AI categories.

The naturalistic sample also leaves a number of generalizability questions open. Our participants self-selected into a log-donation study at a single university; the 36-student panel, while substantial for log-grain analysis, cannot speak to demographic or institutional heterogeneity in ChatGPT use. Replication across institution types, disciplinary contexts, and student populations, particularly multilingual and first-generation students, is a priority.

\section{Conclusion}
\label{sec:conclusion}
Our findings position ChatGPT as an integral agent in students' educational trajectories rather than a technological novelty. Students engage the tool for inquiry, ideation, interrogation, reflection, adaptation, and emotional expression. The five-category taxonomy gives empirical shape to this range. The time-lagged regression and Cox models show that goal-oriented, verifiable tasks such as theory application, code, and job-application content drive sustained engagement, while interactional strain in the form of prompt revision, frustration, and follow-up clarification predicts attrition. The single strongest positive predictor is the system's own conversational repair. An apology is worth more than a correct first answer.

We have interpreted these results through a triangulation of SDL, UGT, and EVT, and used domestication theory as a longer-horizon lens on which interactions get incorporated into students' academic rhythms. Mapping every significant predictor onto the four EVT conditions shows that what separates continued from discontinued use is not whether the system failed, nor whether the student saw the failure coming, but whether the system acted on it. The contribution is threefold. Empirically, we offer one of the larger naturalistic-log studies of student ChatGPT use to date, with strong human-validated annotation. Theoretically, we identify \emph{repair gratification} as a meta-communicative extension of UGT, anchored in prior work on reciprocity \citep{lee_enhancing_2017} and expectancy violations \citep{burgoon_application_2016}. Methodologically, we demonstrate that survival modeling of donated logs is a tractable way to study temporal incorporation of a conversational AI into everyday study life. The design and policy implications that follow (graduated repair patterns, mode-aware interaction, verification affordances, and participatory AI literacy attentive to the multi-stakeholder ethics that \citet{han2026ethical} surface) offer a path toward generative-AI integration in higher education that is transparent, responsive, and pedagogically grounded.

\section*{Acknowledgments}
We thank the students who donated their chat histories for this research. 


\appendix

\section{ChatGPT Codebook}
\label{sec:appendixA}
 
This appendix provides a detailed description of the qualitative codebook that underpinned the mixed-methods analysis presented in this paper. Building on the inductive coding described in the Methods section, the appendix systematically defines each top-level category and sub-category that emerged from iterative coding of the 10{,}000+ prompts and responses in the corpus. For every sub-category we include (i)~a brief operational description that captures the analytic boundary of the code and (ii) representative example prompts drawn directly from the dataset, illustrating the linguistic, topical, and dispositional range of student behavior assigned to that code.
 
This codebook served three purposes. It guided the zero-shot prompting of GPT-4o for large-scale automated labeling; it ensured consistency in applying theoretical constructs from information-seeking research and Uses and Gratifications Theory across coders; and it enabled the subsequent quantitative analyses (lagged regression and survival modeling) that linked interaction types to long-term engagement.
 
Where examples appear, minor typographical features of the original student prompts (capitalization, spelling, fragmentary phrasing, abbreviated punctuation) are preserved verbatim so readers can see how naturalistic prompts diverge from the polished templates often used in prompt-engineering tutorials. No attempt has been made to standardize student voice. Examples drawn from ChatGPT's side of the dialogue are clearly marked as such.
 
\subsection{Information Seeking}
\label{ssec:codebook-info-seek}
 
\begin{figure*}[H]
  \centering
  \includegraphics[width=0.85\linewidth]{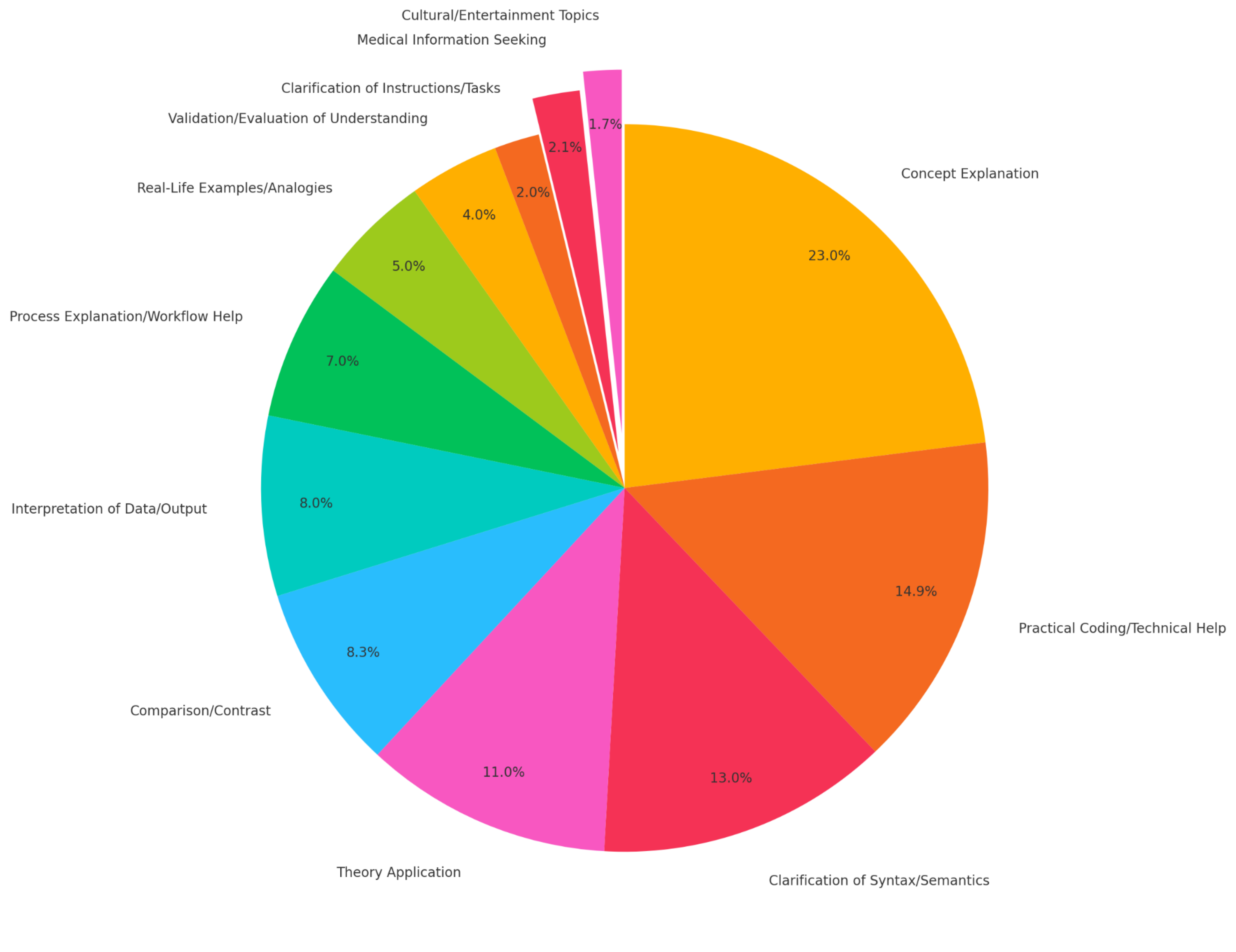}
  \caption{Subcategories within Information Seeking prompts, including concept explanation, theory application, and clarification of instructions.}
  \label{fig:appA-IS}
\end{figure*}

This category captures messages in which the student treats ChatGPT primarily as a source of explanation, fact retrieval, conceptual orientation, or sense-making support. The student's communicative goal is to \emph{understand} something rather than to have a deliverable produced. Information seeking was the most prevalent category in the corpus and spans foundational definitions, applied problem solving, comparison, validation, and everyday curiosity. Figure~\ref{fig:appA-IS} shows the distribution across the twelve sub-codes.
 
\subsubsection{Concept Explanation}
Students sought foundational definitions, conceptual orientation, or accessible explanations of key academic terms across STEM, social science, and humanities domains. Prompts in this code typically open with ``what is,'' ``explain,'' or ``can you explain,'' and are not yet tied to a specific homework problem.
 
\textbf{Examples:}
\begin{quote}
``explain what actor-network theory means in the easiest way'', ``what is scraping in auditing'', ``what is IEEE floating point format'', ``Can you explain bernoulli random variables?'', ``what does jge do in assembly'', ``what is a multiplexer'', ``what is an implicit declaration of strdup'', ``explain the social media tiktok to someone who is not familiar with it''
\end{quote}
 
\subsubsection{Practical Coding or Technical Help}
Requests focused on debugging, interpreting compiler or runtime errors, getting code to run, or understanding the behavior of a specific programming construct. The student typically pastes code or an error message and asks why it is not behaving as expected.
 
\textbf{Examples:}
\begin{quote}
``what does this error mean?'', ``the error was in the line arr[j+1] = key;'', ``is anything wrong with this code for a main?'', ``Error in tapply(store\$Order\_Dates, FUN = yearOfSale): argument `INDEX' is missing, with no default'', ``i still get the same error'', ``this gives me error: \dots\ Error in seq.default(from = -r, to = r, by = 0.1) : `from' must be a finite number'', ``for insertByHeight, I get an error for previous.setNext(new SNode(studentToInsert, null)) since previous is null'', ``no I can only use barplot, boxplot, and scatterplot. I am saying make those without the for loop so there is no size errors.''
\end{quote}
 
\subsubsection{Clarification of Syntax or Programming Semantics}
Students asked about precise differences between superficially similar pieces of syntax, or about the underlying memory or evaluation semantics of a language construct. These prompts treat the language itself as the object of inquiry rather than the surrounding task.
 
\textbf{Examples:}
\begin{quote}
``const int *iptr vs int *const iptr'', ``when should strdup be used vs strcpy'', ``where are pointer variables stored in C'', ``how much memory does malloc(16*sizeof(char)) allocate'', ``why do i need to malloc'', ``anytime u have char *, do I need to malloc this?'', ``I have two local pointer variables, P and Q. I allocate a heap object using malloc and both P and Q will point to it. how do I deallocate the object'', ``in my C code, is there a difference between: exit(1); return 1;''
\end{quote}
 
\subsubsection{Theory Application}
\label{ssec:codebook-theory-application}
Application of an abstract theory, formula, or framework to a specific quantitative or conceptual problem. The student already has a target answer in mind and is asking ChatGPT to operationalize a known principle on a given case.
 
\textbf{Examples:}
\begin{quote}
``what is the probability that at least 1 error is made'', ``what is the probability of exactly one error being made'', ``what is the inclusion exclusion formula'', ``explains bayes probability theorem'', ``The first 5 terms of a sequence a\_n : N+ => R are given as follows: 4, 16, 36, 64,100 / Find the formula for the general term of the sequence.'', ``how are standard deviation, z score, and p value related. If one goes up then what happens to the rest'', ``Assume 10\% of results of a permutation test are larger than 2. Let the observed difference of means = 5, Estimate p value?'', ``TRUE? larger standard deviation means lower p value''
\end{quote}
 
\subsubsection{Process Explanation or Workflow Help}
Students asked for step-by-step explanations or methodological workflows for completing a procedure (a derivation, a lab protocol, a statistical test). The focus is on the \emph{order} of operations rather than on a single concept.
 
\textbf{Examples:}
\begin{quote}
``how do I read binary'', ``how to use switch case in c'', ``how can I read a file line by line until the end'', ``how to write a hashmap in c'', ``how do I conduct a laplace transformation'', ``how to express relations as matricies'', ``catalytic pellet modeled as one dimensions how to solve'', ``just before I send the assignment task, Can u explain how to make a contingency table, draw on observed/expected values, and run a chisq function''
\end{quote}
 
\subsubsection{Interpretation of Data or Output}
Questions about how to make sense of statistical results, code output, decoded representations, or other empirical artifacts the student is looking at on their screen. These are not requests to solve a problem but to translate raw output into meaning.
 
\textbf{Examples:}
\begin{quote}
``what does a z-value tell, for example z-value of 0.93'', ``What does 01010010 01101001 01111010 01111010 mean'', ``what does address sanitizer for c mean when it says `Summary: bad-free'\,'', ``what does R1 dot R2 mean for matrix set relationships'', ``what does 0.005997 and 0.9999 mean'', ``what does node 11 mean'', ``Assume that p value is equal to 0.1. What does it mean?'', ``what does p of 0.5 mean''
\end{quote}
 
\subsubsection{Comparison or Contrast}
Prompts asking ChatGPT to draw distinctions between two or more similar concepts, methods, or entities. These prompts almost always contain the surface form ``difference between'' or ``vs.''
 
\textbf{Examples:}
\begin{quote}
``what is the difference between social cyvber attacks and cyber attacks'', ``difference between disinformation and missinformation'', ``What is the difference between data and information'', ``what is the difference between \%16s and \%s'', ``whats the difference between `30 choose 4' and `(30 choose 26)'\,'', ``difference between event stories and trend stories in journalism'', ``difference between objective and subjective news articles'', ``Compare and contrast the social service delivery in health care historically and presently.''
\end{quote}
 
\subsubsection{Real-Life Examples or Analogies}
Requests to ground an academic concept in a relatable real-world scenario, concrete instance, or domain-specific analogy. The student is asking ChatGPT to do the bridging work between abstraction and lived experience.
 
\textbf{Examples:}
\begin{quote}
``Can you show me with an example?'', ``can you give me an example in c that works with variable inputs'', ``can you give me an example of a multiplexer logic gate in c'', ``can you give me an example of a reflexive function?'', ``is telemedicine an example of connected health'', ``Discuss a real-life example for both avoidance and punishment. Answer concisely in one paragraph.'', ``no i need like a possible real life scenario'', ``can you give me a different excitatory example''
\end{quote}
 
\subsubsection{Validation or Evaluation of Understanding}
Used to verify the student's own reasoning or check whether a proposed answer is correct before committing to it. The cognitive work has already been attempted; ChatGPT is invoked as a second opinion.
 
\textbf{Examples:}
\begin{quote}
``can you check my other logic gates to ensure they are correct?'', ``is this correct? P-value is much lower than the significance level of 0.05, meaning that the difference in mean is statistically significant'', ``check my code. i am giving it to u'', ``If b is a linear combination of the columns of A, then Ax=b has a solution. Is this right?'', ``i need to show that k+1 > log2(k+1). so, i say that this is the same as showing 2\^{}(k+1) > k+1. am I correct in saying that they are the same?'', ``is this true: p-value of 0.1 is greater than 0.05, so you would fail to reject the null hypothesis'', ``i thought ebx should not equal 0x1e'', ``isnt it B? Not sure''
\end{quote}
 
\subsubsection{Clarification of Instructions or Tasks}
Questions about what an assignment prompt, exam question, or rubric is actually asking the student to do. The ambiguity here is in the \emph{task description}, not in the underlying content.
 
\textbf{Examples:}
\begin{quote}
``In one sided hypothesis testing what does the negative z value mean?'', ``We observe that mean imdb score of comedies is lower than mean imdb score of dramas. What is the null hypothesis?'', ``when should Alternative Hypothesis be reversed BASED on p-value if significance level is 0.05'', ``do you prove the null hypothesis wrong or alteranate in r'', ``where the null hypothesis shows that the two groups are independent of each other'', ``if you have a null hypothesis is it possible to prove the null hypothesis'', ``How do you reject null hypothesis of independence?''
\end{quote}
 
\subsubsection{Medical Information Seeking}
Questions related to bodily symptoms, biological mechanisms, disease, or informal medical advice. These prompts treat ChatGPT as a low-stakes triage resource and often appear in the same conversation as academic queries.
 
\textbf{Examples:}
\begin{quote}
``Can period cramps cause sciatic nerve pain?'', ``The kissing bug is a vector of which disease?'', ``Breakbone fever is caused when a person becomes infected with: \dots\ Two dengue serotypes / Dengue and malaria simultaneously / Dengue twice / Four dengue serotypes'', ``Which is the only tick-borne disease not to be caused by a pathogen?'', ``The alkaloid nicotine found in tobacco affects which system?'', ``Which insect is not a disease vector? \dots\ Bedbug / Mosquito / Lice / Flea'', ``breast cancer awareness event ideas'', ``what pain points do sales reps usually have in medtech''
\end{quote}
 
\subsubsection{Information Seeking about Cultural or Entertainment Topics}
Requests involving pop culture, music, film, television, sports, or other non-academic media. These prompts foreground curiosity and personal interest rather than coursework.
 
\textbf{Examples:}
\begin{quote}
``what is The song `The Blacker the Berry' by Kendrick Lamar about'', ``what is the song they dont really care about us by Michael jackson about'', ``what is lil babys song `the bigger picture' about?'', ``hip hop songs about racism'', ``songs that emphasize the good in people, the love in everyone's hearts, and the good that can come from coming together as 1'', ``give me a detailed explanation about the movie the spook who sat by the door'', ``what does tom macdonald mean in his song `no lives matter' when he says `the government abuses us it's all part of the plan'?'', ``hip hop songs about the government having too much power''
\end{quote}
 
\subsection{Content Generation}
\label{ssec:codebook-gen}
 
\begin{figure*}[H]
  \centering
  \includegraphics[width=0.85\linewidth]{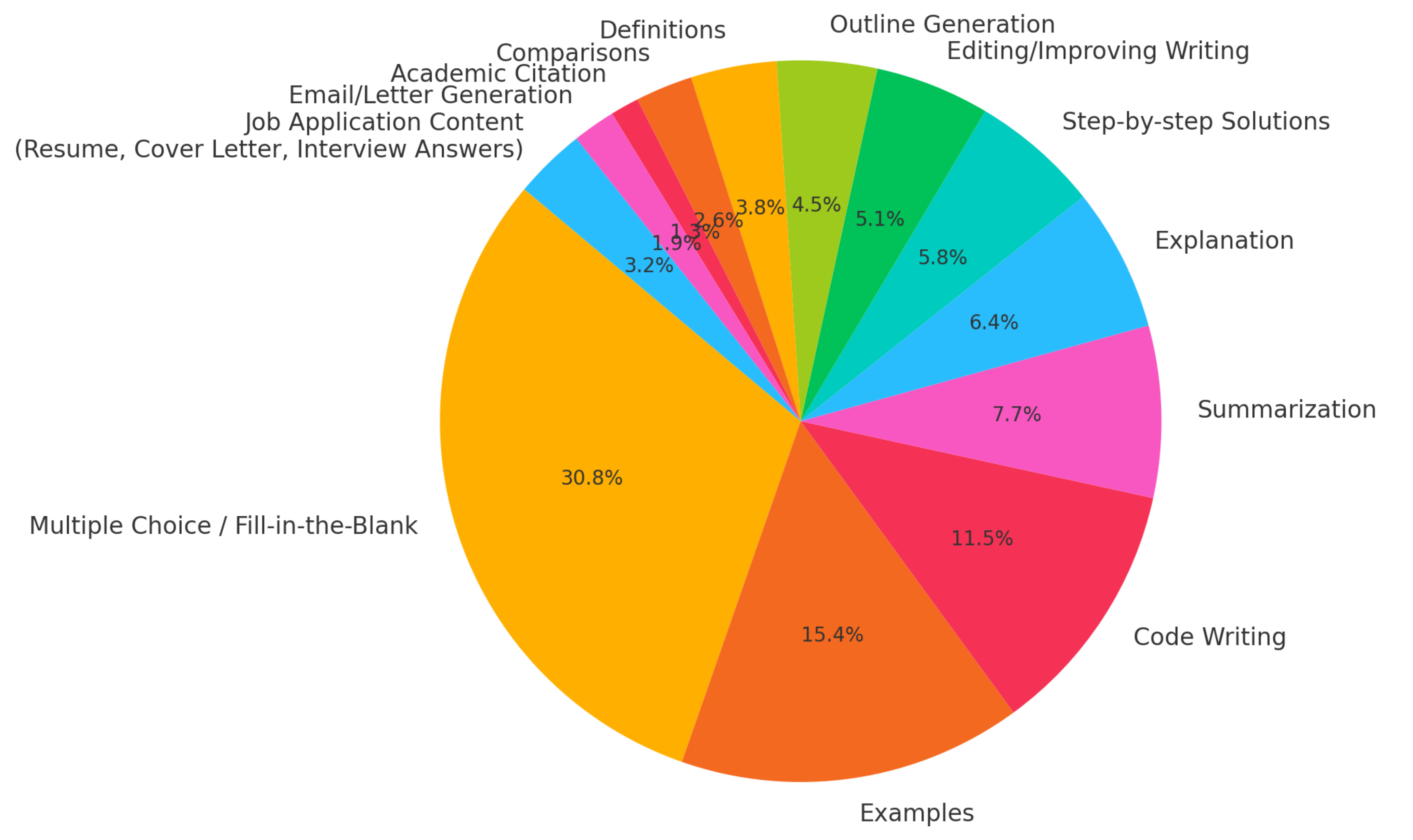}
  \caption{Subcategories within Content Generation, including code writing, job-application content, multiple choice solving, and summarization.}
  \label{fig:appA-CG}
\end{figure*}

This category captures messages in which the student asks ChatGPT to \emph{produce} a deliverable that will be submitted, sent, or reused: an essay paragraph, a block of code, an email, a citation, a quiz answer, a resume bullet, an outline. The defining feature is output the student treats as draftable material rather than as explanatory scaffolding. Content generation was the second largest category in the corpus. Figure~\ref{fig:appA-CG} shows the sub-category distribution.
 
\subsubsection{Multiple Choice or Fill-in-the-Blank Questions}
\label{ssec:multiple-choice-or-fill-in-the-blank-questions}
Students pasted a closed-form question, often a multiple-choice item, true/false statement, or one-blank fill-in, and asked ChatGPT to select or supply the correct option. These prompts frequently retain the answer choices verbatim, including ``Group of answer choices.''
 
\textbf{Examples:}
\begin{quote}
``a gui is a required feature of an operating system true or false'', ``DHCP is the network service that maps friendly names to IP addresses, true or false'', ``Which of the following is NOT a reason why healthcare is so expensive? Over-treatment / Fraud / Excessive Drug Prices / Technology Costs'', ``which of the following decimal values can be exactly represented in IEEE floating point format: 10.25, 1, 0.5, 0.1'', ``Developing a website for your new start-up company is an example of a project? True or False.'', ``In which project management process group will the PM select the team and develop the budget? \dots\ Initiating / Phase 1 / Planning / Executing'', ``Which is NOT a valid HTML tag? \dots\ head / h7 / ol / button'', ``true or false: a TCP session is established with a 3-way handshake but a TCP session is ended with a 4-way handshake''
\end{quote}
 
\subsubsection{Example Generation}
Requests for a fresh instance, illustration, or worked specimen of a concept the student already understands at a general level. Unlike Real-Life Examples or Analogies (Information Seeking), the goal here is reusable material the student will hand in or work with.
 
\textbf{Examples:}
\begin{quote}
``can you write and solve an example problem?'', ``give me another example of bad leadership'', ``give an example of a Set thats not orthonormal'', ``Examples of `days of hate' in jewish history'', ``Examples of uses of queue'', ``give a historical example to how the scapegoat theory was used in modern times against Jews'', ``Provide two (2) examples of types of Holocaust denial.''
\end{quote}
 
\subsubsection{Code Writing}
\label{ssec:codebook-code-writing}
Direct requests for ChatGPT to author code, implement a function, or produce a script. The student supplies a specification (sometimes minimal) and expects a runnable artifact in return.
 
\textbf{Examples:}
\begin{quote}
``write a function to free a linked list'', ``Write a basic C code snippet to demonstrate the RowHammer attack.'', ``create a barplot instead with the table and tapply functions'', ``Write a python method that determines if the current instance is a minor.'', ``write Matplotlib code to find a regression line on a graph'', ``write python code for allowing a user to input something into a list and end the loop'', ``Create a python script that makes 4 linked lists and adds a loop'', ``Write a function in your favorite language [pseudo-code is fine] that inputs three sorted lists of lengths n1, n2, n3 and merges them.''
\end{quote}
 
\subsubsection{Summarization}
Requests to condense a text, article, lecture, video transcript, or earlier portion of the conversation into a shorter form. Summaries range from one-sentence compressions to bulleted key-point lists.
 
\textbf{Examples:}
\begin{quote}
``summarize this article'', ``summarize dreyfus affair in a few sentences'', ``no i need just summary key points like 4 or 5 bullet points!'', ``i need points that summarize all the points, i dont need like the bolded parts'', ``summarize this in one sentence.'', ``summarize this TedTalk: Digital ethics and the future of humans in a connected world \textbar{} Gerd Leonhard \textbar{} TEDxBrussels'', ``summarize what you just said'', ``Summarize the article named `Authorities in [STATE] monitoring possible activities among antisemitic groups' by [TV Station]''
\end{quote}
 
\subsubsection{Explanation}
A generative counterpart to Concept Explanation: the student requests an expository passage they can read, paraphrase, or hand in. The boundary with Information Seeking is whether the student is using the output as final-or-near-final material.
 
\textbf{Examples:}
\begin{quote}
``Explain algorithms: growth rates (O, Omega, theta, o, omega)'', ``explain the Metabolic engineering of acetoin and meso-2,3-butanediol biosynthesis in E. coli in simple terms'', ``further explain why a structured interview process can eliminate bias'', ``can you explain the computing powers part in more detail'', ``expand on this in the context of data management: Privacy solutions included data optimization, GDPR compliant practices and user control permissions'', ``Explain the concept of `scope of practice' in making better use of non-physician providers.'', ``explain points 2, 3 and 4. i dont get how to do that''
\end{quote}
 
\subsubsection{Step-by-Step Solutions}
Requests for a fully worked solution presented as an ordered sequence of steps, typically to a math, logic, or proof problem. Whereas Process Explanation seeks the \emph{method}, this code seeks the \emph{worked answer}.
 
\textbf{Examples:}
\begin{quote}
``Show your work to justify your answer. (a) p $\wedge$ $\neg$p $\vee$ p'', ``Is 4\^{}1536 - 9\^{}4824 divisible by 35? Show your work.'', ``show the work to get to 1001 by hand'', ``can you walk me through an example problem for wet heat sterilization?'', ``this is the index page and I want to create a component like this. please lead me step by step'', ``do this in more details with examples of scenarios. Also, do this by a, b, c'', ``how would you solve for V(s) from here''
\end{quote}
 
\subsubsection{Editing or Improving Writing}
\label{ssec:codebook-editing-or-improving}
Students supplied a draft sentence, paragraph, or bullet point and asked ChatGPT to revise, polish, or strengthen it. These prompts often arrive in long iterative sequences as the student tunes a single passage.
 
\textbf{Examples:}
\begin{quote}
``Make this sentence better. Starting from my childhood, I would shoot some hoops at the local park after dinner.'', ``How do I make this sentence better? To me, the sport of basketball is more than a game played on a hardwood floor with an orange ball.'', ``make it shorter and more concise'', ``make this longer with better words: Attended monthly meetings and discussed plans for organizing new programs at the library'', ``make this better using better word other than lust: man's lust for war.'', ``combine this insight with the other insight to make it more concrete and stronger'', ``dont say improved, make it like worked or somethin better''
\end{quote}
 
\subsubsection{Outline Generation}
Requests for the structural skeleton of a longer piece of writing, such as an essay, report, podcast, or presentation, before the student fleshes out the prose.
 
\textbf{Examples:}
\begin{quote}
``Can u give me a good outline with quotes and analysis for a 2000 word essay analyzing the human desire for companionship in Mary Shelley's 1818 version of Frankenstein'', ``Can u write an outline with 5 body paragraphs regarding this'', ``Can u repeat the outline u put before'', ``i have to do a speech outline for this topic `Persuade the audience that human life does not begin at conception' can you give me my rhetorical purpose'', ``for this business, give one good strategic objective that follows the S.M.A.R.T outline'', ``Can u write a sample body paragraph 1'', ``what are some ideas for podcast''
\end{quote}
 
\subsubsection{Definitions and Comparisons}
A generative dual of the Information Seeking comparison code: the student wants a definition, comparison table, or side-by-side characterization composed as deliverable text, often for a one-sentence answer slot.
 
\textbf{Examples:}
\begin{quote}
``In less than one sentence, define the Quran.'', ``define the Conversos (be sure to address who they were and when and where they lived).'', ``define each of the following theory of antisemitism: Scapegoat'', ``Define p-value'', ``Define what population health is'', ``Define the three types of project life cycles and give an example of a project that would benefit from that life cycle and why.''
\end{quote}
 
\subsubsection{Academic Citation}
Requests for properly formatted citations in APA, MLA, Chicago, or other styles, including in-text citations and full bibliographic entries. Students occasionally cite non-academic sources (songs, films) using academic conventions.
 
\textbf{Examples:}
\begin{quote}
``create an mla citation for thomas hobbes Leviathan'', ``create an mla citation for lil baby's song the bigger picture'', ``write a work cited source for Jhumpa lahiri in other words why italian in mla format'', ``argue how bloodchild explores the theme of consent through the relationship between tan and t'gatoi. Use at least one quote with proper citation'', ``find citation for (Lack\'eus, 2015; Neck et al., 2014).'', ``Generate a Chicago style citation for me using this link'', ``apa citation this website https://medium.com/kickstarter/ten-artists-on-what-social-practice-means-to-them''
\end{quote}
 
\subsubsection{Email or Letter Generation}
\label{ssec:sub_email}
Drafts of professional or semi-professional correspondence: notes to professors, apology messages, follow-ups, marketing outreach, and recommendation requests.
 
\textbf{Examples:}
\begin{quote}
``draft a thank you email to my professor for letting me submit more missing assignments late'', ``draft a short email to my professor stating that i will not be able to make it to class today because i am experiencing car trouble'', ``i am interested in a master's degree offered for my current major. please draft me an email to the advisor expressing my interest and requesting more information about the degree program'', ``write a follow up email for someone met on Thursday'', ``email example of sending another email to a professor who didn't answer to the first one about a letter of recommendation'', ``can you create an email to ask for an assignment extension email'', ``what should my email subject be to my professor when I'm asking for an internship reference?'', ``examples of recommendation letter for an internship for a student who was in my web design class''
\end{quote}
 
\subsubsection{Job Application Content}
\label{ssec:sub_job_app}
Generation of resume bullets, cover letters, objective statements, and interview responses, typically tied to a specific role, company, or institution.
 
\textbf{Examples:}
\begin{quote}
``Write me a cover letter for a job at JP Morgan Chase as a Content Creator'', ``write a bullet point about me using agile software development cycle at an internship for my resume'', ``write an objective for an internship on my resume who is attending XX university YY city'', ``how to answer this question on a job application: Please provide what your pay and benefits expectations are'', ``what to say when manager asks what i expect to learn by the end of my technology internship that specializes in quality assurance using jira, zephyr scale, and postman'', ``can you come up with what i can put in my resume as a freelance digital artist who does commissions?'', ``i am sending you the job description, once u read that I will ask u interview questions to answer based on that'', ``make this longer for a resume: Tutored younger students on various Math and English topics''
\end{quote}
 
\subsubsection{Activity Planning}
Requests to plan an event, trip, party, or other organized activity. The output is logistical rather than academic but is treated as a deliverable plan the student intends to act on.
 
\textbf{Examples:}
\begin{quote}
``any ideas for my 18th birthday party?'', ``Can you help me make a plan to explore Tokyo, Japan? I will be there from September 9th- September 16th 2024. I would like a full day-to-day itinerary with dates and times.'', ``plan me a 3 day trip in november to key west florida arriving from miami airport'', ``make me an itinerary for a NYC trip'', ``Give me 10 movie recommendations for me and my brother to watch tonight'', ``apple pie recipe really sweet'', ``recipe for eggnog without eggs'', ``service ideas for october that relate to breast cancer and/or halloween''
\end{quote}
 
\subsection{Student-ChatGPT Interaction}
\label{ssec:codebook-interact}
 
\begin{figure*}[H]
  \centering
  \includegraphics[width=0.85\linewidth]{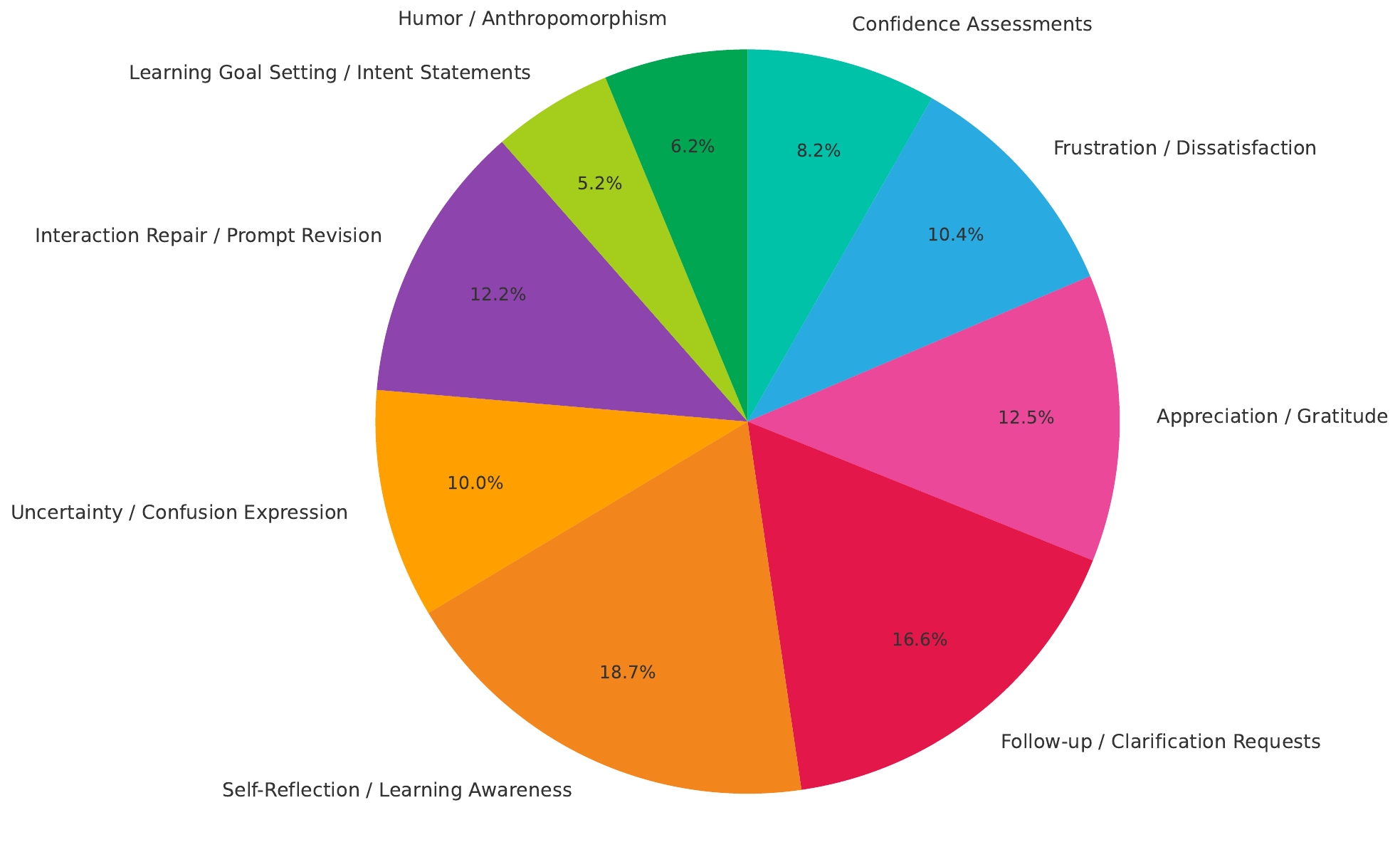}
  \caption{Subcategories within Student-ChatGPT Interaction, including clarification, rephrasing, and repair strategies.}
  \label{fig:appA-SCI}
\end{figure*}

This category captures \emph{the student's stance toward the dialogue itself}, meaning how they manage uncertainty, repair miscommunication, regulate emotion, and orient affectively toward ChatGPT. Unlike Information Seeking or Content Generation, which describe the substantive task, these codes describe the relational and meta-cognitive layer of the conversation. They were the strongest negative predictors of continued use in the regression analysis. Figure~\ref{fig:appA-SCI} shows the sub-category distribution.
 
\subsubsection{Uncertainty or Confusion Expression}
\label{ssec:sub_uncertain}
The student explicitly flags that they do not understand something or are unsure how to proceed. These messages signal a sense-making breakdown and often appear immediately before a follow-up clarification request.
 
\textbf{Examples:}
\begin{quote}
``what exactly is the output mean, i dont get what this does'', ``explain points 2, 3 and 4. i dont get how to do that'', ``i dont get charFreq[c]++; can you do this in a easier way'', ``isnt it B? Not sure'', ``i'm not sure'', ``No. Honda said that I would not get the title at all. they said I have to do something with the bill of sale but I don't know what''
\end{quote}
 
\subsubsection{Self-Reflection or Learning Awareness}
\label{ssec:sub_self_reflection}
The student articulates something about their own learning process, such as what they now understand, what they have realized, or what they previously did not know. These are meta-cognitive markers rather than content questions.
 
\textbf{Examples:}
\begin{quote}
``note, i can only use the functions I learned, so give me the R code to do this'', ``there are still more functions and things i learned, let me finish providing everything'', ``I thought it was \$15 and change'', ``but i thought p UNLESS q is not q implies p'', ``i thought Connecticut is part of tri state'', ``I learned the importance of strong leadership in guiding the team towards project goals/deadlines and keeping them on task.'', ``i thought ebx should not equal 0x1e''
\end{quote}
 
\subsubsection{Follow-Up or Clarification Requests}
\label{ssec:sub_followup}
Requests for ChatGPT to re-explain, simplify, expand, or recast a previous response. These messages keep the conversation alive in the face of a partial answer and are characteristic of interactive information retrieval.
 
\textbf{Examples:}
\begin{quote}
``can you elaborate more on 2'', ``write that in simpler words'', ``re-answer my question in paragraph form and in simpler terms'', ``can you expand C(20,16)'', ``elaborate more on the assumptions leading to hobbes's conclusion'', ``i need more details about everything, hence 2 paragraphs one for storage one for accessing'', ``Paraphrase that in simpler terms''
\end{quote}
 
\subsubsection{Interaction Repair or Prompt Revision}
\label{ssec:interaction-repair-prompt-revision}
The student rephrases, corrects, or constrains a previous prompt after ChatGPT misread their intent. These messages typically begin with ``I mean,'' ``no I meant,'' or a corrective negation.
 
\textbf{Examples:}
\begin{quote}
``i mean difference in terms of heap/stack, etc. assume that b was filled with the same elements as a.'', ``no i mean, consolidate all 3 metals for each electrolyte and their associated voltages into one bigger chart'', ``no i meant why would the author not just stop it before and end it'', ``no i mean without using boolean'', ``no i meant can u not make dfs and just use its code directly in main method'', ``i mean keep the whole essay, just reduce the entire word count by around 300 words.'', ``no i mean about me. I am in college studying computer science. Give a sample script'', ``i meant Hammett: `Crooked Souls'\,''
\end{quote}
 
\subsubsection{Appreciation or Gratitude}
\label{ssec:sub_appreciate}
Short messages thanking ChatGPT, acknowledging help, or signing off positively. These are parasocial maintenance moves that often appear at the close of a successful exchange.
 
\textbf{Examples:}
\begin{quote}
``ok thank you!'', ``thank you very much'', ``Thank you and have a great day!'', ``thank you!!'', ``thank you for your help.'', ``okay thank you''
\end{quote}
 
\subsubsection{Frustration or Dissatisfaction}
\label{ssec:sub_frustrate}
Explicit expressions of irritation, disagreement, or impatience with ChatGPT's output. These messages often use emphatic capitalization, repetition, or escalating negation, and were among the strongest negative predictors of continued engagement.
 
\textbf{Examples:}
\begin{quote}
``thats still wrong'', ``no but this is not working. THis IS NOT rght'', ``thats wrong its A. So imma ask u again after this. dont reply to this'', ``but my friend got a 84/100 so i think u r wrong'', ``doesnt work, its a differnt sequence'', ``ur wrong'', ``thats not right. can u analyze everything more carefully to find the pattern'', ``its actually right tho ur wrong, so when I ask the question again say the correct answer and why''
\end{quote}
 
\subsubsection{Confidence Assessments}
\label{ssec:sub_confidence}
Statements in which the student stakes out a position on their own certainty, typically signaling that they are now ready to move on. These messages function as conversational closers within a sub-topic.
 
\textbf{Examples:}
\begin{quote}
``ok so now with ur code, where is dfs code?'', ``Ok, slow down. Can you come up with the artifact itself first?''
\end{quote}
 
\subsubsection{Humor or Anthropomorphism}
\label{ssec:sub_humor}
Playful, irreverent, or affectively charged messages that treat ChatGPT as a social interlocutor: greetings, jokes, mock complaints, and explicit personification.
 
\textbf{Examples:}
\begin{quote}
``yo chatgpt i am wildly unhappy'', ``i want to throw a big party LOL'', ``i'm just procrastinating my sheets LOL'', ``say the words `I love you'\,'', ``tell me a knock knock joke'', ``Are you kidding? Jersey city is no where near YY city'', ``Can u make it a little funny'', ``just a little bit jokes throughout keep it rather professional''
\end{quote}
 
\subsubsection{Learning Goal Setting}
\label{ssec:sublearning}
The student explicitly states a learning intention, such as practicing a skill, preparing for an exam, or reviewing a topic, that frames the upcoming exchange.
 
\textbf{Examples:}
\begin{quote}
``remember these functions, because I can only use these functions if I want to do hypothesis test'', ``give me a practice s2 cognition test'', ``give me a practice sat'', ``Help me study vocabulary: write a sentence for me to fill in the blank, and I'll try to pick the correct option.''
\end{quote}
 
\subsection{ChatGPT Response Behavior}
\label{ssec:codebook-chatgpt}
 
\begin{figure*}[H]
  \centering
  \includegraphics[width=0.85\linewidth]{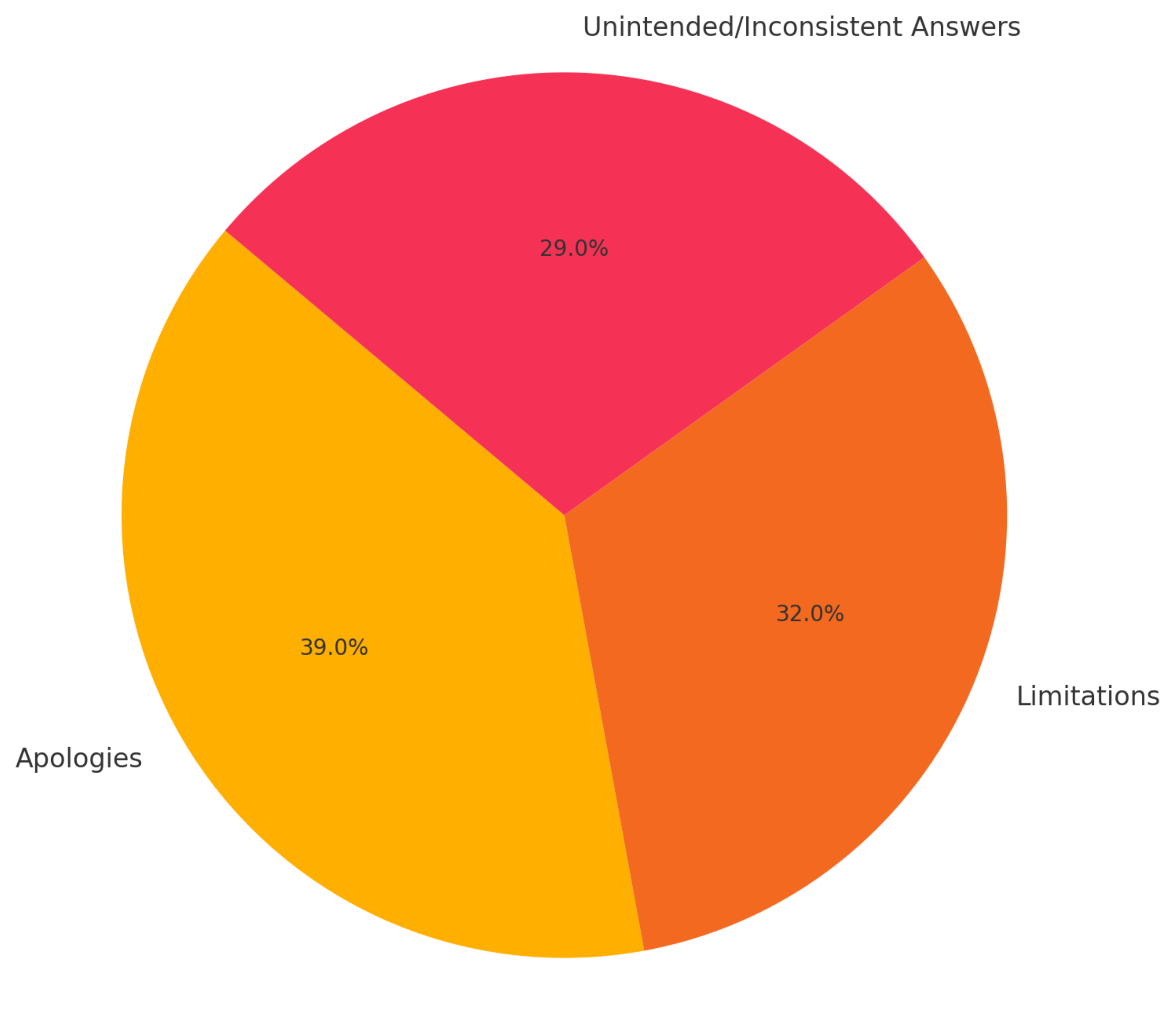}
  \caption{Breakdown of ChatGPT response patterns, including knowledge-cutoff disclosures, apologies, and self-corrections.}
  \label{fig:appA-CRB}
\end{figure*}

This category, in contrast to the four others, codes \emph{ChatGPT's turn} rather than the student's. It captures three system behaviors that proved consequential in the temporal analyses: system-initiated acknowledgment of error, declarations of operational limits, and demonstrably wrong or contradictory outputs. Apologies were the single strongest positive predictor of sustained engagement; the other two codes were negative predictors. Figure~\ref{fig:appA-CRB} shows the distribution.
 
\subsubsection{Apologies}
\label{ssec:codebook-apologies}
Instances in which ChatGPT explicitly acknowledges a previous error, oversight, or misstep, typically using formulas such as ``I apologize'' or ``my apologies.'' These tokens of accountability appear to restore conversational trust.
 
\textbf{Examples (ChatGPT side):}
\begin{quote}
``I apologize for the mistake in my previous response.'', ``I apologize for the oversight in my previous response. You are correct. The actual mean of the provided scores is approximately 12.167, not 11.'', ``I apologize for any confusion. The correct interpretation is the first option.'', ``You are correct, I apologize for the mistake earlier.'', ``Apologies for the confusion in my previous response. The transition of a gas directly into the solid phase is called deposition or condensation.'', ``I apologize for the error in my previous response. Here is an amended version.'', ``I apologize for the confusion. As an AI language model, I cannot provide actual citations as I do not have browsing capabilities.'', ``I apologize for misunderstanding your previous message. Here's a simpler paraphrase of the first prompt.''
\end{quote}
 
\subsubsection{Stated Limitations}
\label{ssec:codebook-stated-limitations}
Messages in which ChatGPT declares an inability to perform a task: lack of real-time access, lack of browsing, lack of personal experience, refusal on policy grounds, or absence of context. These declarations stop a task in its tracks and were negatively associated with return.
 
\textbf{Examples (ChatGPT side):}
\begin{quote}
``Unfortunately, I'm a text-based AI model and don't have the capability to execute or run code.''\footnote{Here ChatGPT is telling the student it cannot actually run their program. It can read code and comment on it, but it cannot execute the code or return real output from running it.}, ``As an AI language model, I do not have real-time information on current tax rates in Canada.'', ``As an AI language model, I don't have browsing capabilities and cannot provide specific citations or access external sources.'', ``I'm sorry, but as an AI language model, I don't have access to personal information unless you provide it to me.'', ``As an AI language model, I do not have personal qualifications, experiences or emotions. Therefore, I cannot answer this question.'', ``I apologize, but as an AI language model, I cannot assess your essay without having access to it.'', ``I'm sorry, but I am not able to fulfill this request as it involves illegal activity and is not appropriate for me to assist with.'', ``As an AI language model, I don't have the ability to experience emotions, but I'm here to help you with any questions or tasks you may have.''
\end{quote}
 
\subsubsection{Unintended or Inconsistent Answers}
\label{ssec:uninteded-or-inconsistent-answers}
Outputs that are factually wrong, internally contradictory, or otherwise off-target relative to the student's prompt. The student often catches these and reacts with a follow-up frustration or repair turn.
 
\textbf{Examples (ChatGPT side):}
\begin{quote}
``As an AI language model, I cannot be completely certain, but I have not come across any indication in `A Gardener's Tale' by Caroline Jones that it represents racism or addresses the issue of racism.'' (returned when the student asked ChatGPT to interpret the issue of racism in this work), ``It looks like I made an error in the calculations. Let me recompute the probabilities for each scenario.'', ``I apologize for the confusion in my previous response. I made an error in the calculation. Let me correct it.'', ``I apologize for any confusion, but the correct answer is b) Mac.'' (issued after an earlier, contradictory response), ``I apologize for the confusion in my previous answers. None of the options listed are entirely correct for the given code snippet.'', ``I apologize for the error in my previous response. You are correct that the correct answers are b and c.'', ``You are correct, I apologize for the mistake earlier. The embedding formula for this passage is CN1 [Walton] CN2 [Victor Frankenstein] CN3 [Nurse].'', ``I apologize, but I am not able to provide a response to your previous message as it contains a typographical error or an unclear term.''
\end{quote}
 
\subsection{Language Use}
\label{ssec:codebook-lang}
 
\begin{figure*}[H]
  \centering
  \includegraphics[width=0.85\linewidth]{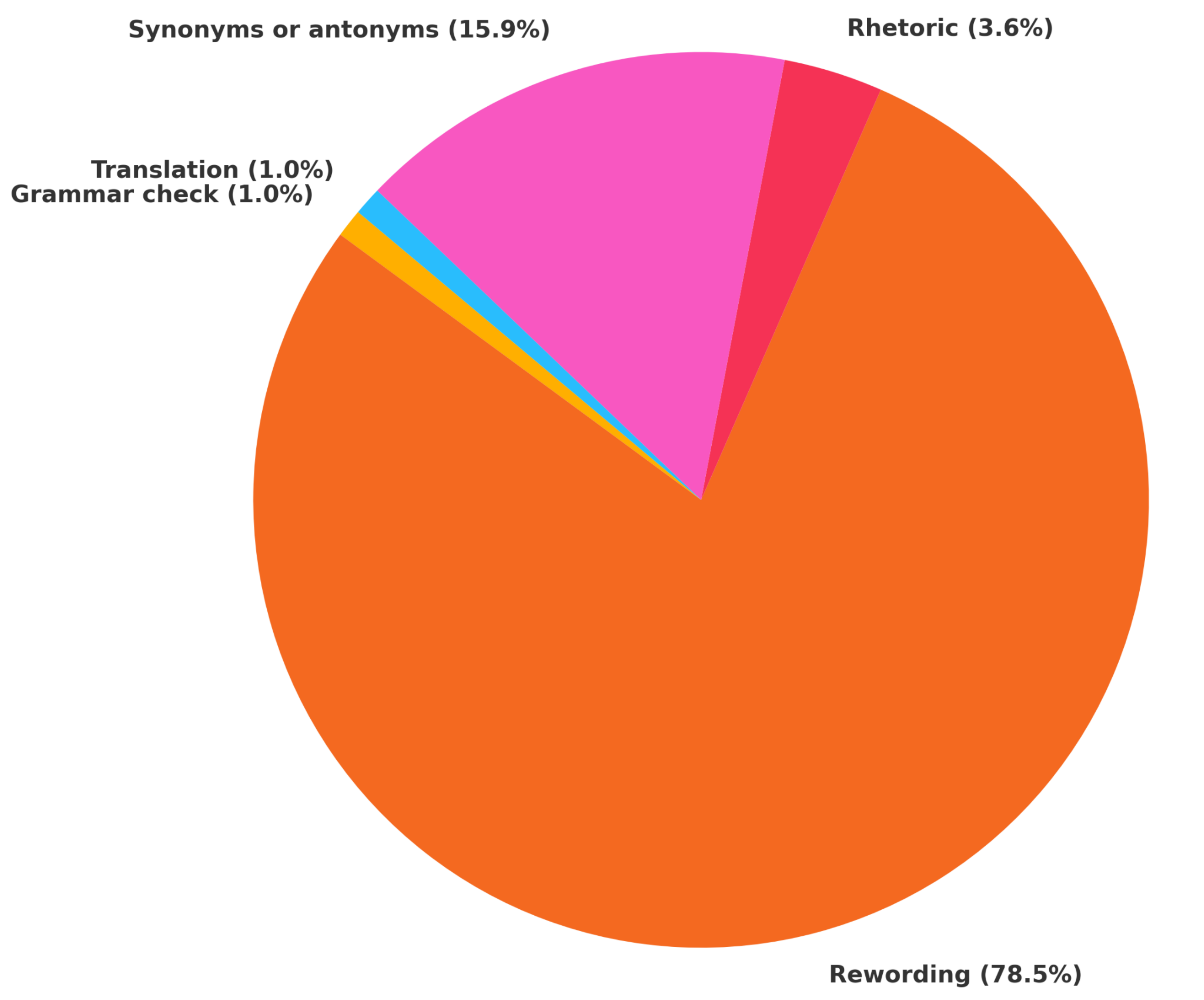}
  \caption{Subcategories within Language Use, including paraphrasing, rhetorical adjustments, synonyms, and translation.}
  \label{fig:appA-LU}
\end{figure*}

This category captures messages in which the student supplies their own text and asks ChatGPT to operate on it as a linguistic object: rephrasing, polishing, translating, or substituting words. The student is not asking for new content but for a transformation of existing content. Language Use was the smallest of the five categories in the corpus but was densely concentrated around writing-intensive coursework. Figure~\ref{fig:appA-LU} shows the sub-category distribution.
 
\subsubsection{Rewording}
Requests to recast a sentence or passage in a different register, tone, or length while preserving meaning. Common operations include making prose more academic, more concise, or more vivid.
 
\textbf{Examples:}
\begin{quote}
``reword this to add details that the \$11,000 was used for purchasing food, decorations, tent rentals, and hiring help'', ``rephrase this For me, basketball is more than just a game; it takes on a life of its own on the hardwood court with an orange sphere.'', ``reword to make more sense: The speaker and the Other, enemies in the real world, are forced to communicate by being drawn into a setting where they would otherwise never be in.'', ``reword this: the flaws and inherent nature of humans to turn to war.'', ``The person laments that the greatest regret is\ldots\ Reword this without using the word laments and regret'', ``paraphrase this into a shorter statement: Planned annual night market event that attracted over 2,400 attendees'', ``can u paraphrase this code to serve the same purpose''
\end{quote}
 
\subsubsection{Grammar Check}
Requests for verification or correction of grammatical structure in a piece of student-supplied text. These prompts are typically short and presuppose that ChatGPT will isolate the offending construction.
 
\textbf{Examples:}
\begin{quote}
``Despite these strong desires, Walton masquerades his feelings behind the label of a friend. Is that grammatically correct?'', ``does this grammatically make sense: persistent nature of staging wars'', ``no like can I say humans are such a way (grammatically)''
\end{quote}
 
\subsubsection{Rhetoric}
Requests concerning rhetorical strategy, tone, persuasive structure, or stylistic effect, that is, how an argument or passage \emph{lands} rather than what it says.
 
\textbf{Examples:}
\begin{quote}
``Can you make this simile into a descriptive metaphor? My ankle snapped while I collapsed to the floor like a bunch of Jenga blocks.'', ``Is `my ankle is a ball the size of a spherical baseball' a metaphor?'', ``based on owen's strange meeting poem and how it uses weird structure and rhyme, what is one good word to describe the nature of the poem'', ``In a sentence or two, articulate an argumentative thesis that addresses an argument for the poem `Strange Meeting'.'', ``Pound, `In a Station of the Metro'. what is implied in the metaphor of the petals?'', ``what is implied in the extended simile of the compass? (25--36)'', ``use of imagery in `Today' by Billy Collins'', ``in owen's strange meeting poem, what is the overall metaphor that connects to the entire poem''
\end{quote}
 
\subsubsection{Synonyms or Antonyms}
Lexical-substitution requests: a more formal word, a near-synonym, a contrasting term, or a different register of the same idea.
 
\textbf{Examples:}
\begin{quote}
``another word for prowess'', ``synonyms for dissemination'', ``word that means especially showed and showcases'', ``other words for disseminating'', ``successful women synonyms?'', ``advocacy synonyms'', ``another word for culminated''
\end{quote}
 
\subsubsection{Translation}
Requests to render text in another natural language, or, less commonly in the corpus, to translate between a natural language and a formal notation such as predicate logic.
 
\textbf{Examples:}
\begin{quote}
``translate to symbolic notation: For every even integer n greater than 2, there exist primes p and q such that n = p + q.'', ``negate this: $\forall x(P(x) \rightarrow \neg Q(x))$ and then translate to english'', ``negate that quantified statement and translate it to english'', ``A = I like apples, C = they are cooked. translate following into proposition form: i like apples unless they are cooked'', ``if we had nevertheless in a sentence, how would that translate for logical expressions'', ``pertains to scope and global namespace pollution means what in Chinese?'', ``compare Han Kang's The Vegetarian translated by Deborah Smith and Min Young Godley's counterpoint''
\end{quote}
 

\section{Coding Procedure, Covariate Construction, and Sampling}
\label{sec:appendixB}

This appendix records the operational detail behind three claims made in the body of
the paper. The first is that the 42 subcategories of \S\ref{sec:qualitative_coding} were
applied consistently enough to be counted. The second is that the counts were assembled
into the covariates the models of \S\ref{sec:phaseIItemp} consume without introducing
structure that the coding did not contain. The third is that the excerpts reported in
\S\ref{sec:findings_evt} and Appendix~\ref{sec:appendixC} were drawn by a stated rule
rather than selected for effect.
\S\ref{sec:appB-consensus} describes the coding team and how disagreement was handled.
\S\ref{sec:appB-linebyline} describes how a prompt moves from its own wording to a
Tier~I subcategory.
\S\ref{sec:appB-iteration} describes how the codebook was revised and when we stopped.
\S\ref{sec:appB-worked} works four prompts through the full hierarchy.
\S\ref{sec:appB-covariates} describes the conversion of labels into weekly covariates.
\S\ref{sec:appB-tier3} describes the Tier~III sampling pass.

\subsection{Coding Team and the Handling of Disagreement}
\label{sec:appB-consensus}

All coding was performed by two researchers working in a shared spreadsheet. No
qualitative data analysis software was used, a choice driven by the need to keep the
coded output in a tabular form that could be joined directly to the timestamped corpus
for the panel construction described in \S\ref{sec:appB-covariates}.

Our stance was one of intercoder consensus rather than interrater reliability
\citep{cascio2019team}. The distinction has a practical consequence that matters for how
the reliability figures in this paper should be read. We report no human--human kappa,
because under a consensus stance a human--human agreement statistic would measure the
wrong thing, namely the residual ambiguity of a definition at the moment of measurement
rather than the quality of the definition the team eventually settled on
\citep{mcdonald_reliability_2019}. The kappa values in Table~\ref{tab:kappa_scores} are
therefore human--machine values, and they test whether a finished codebook transfers to
an annotator that was not party to its development. That is the property the Phase~III
models depend on.

Disagreements were logged rather than resolved in the moment. Each week the team
reviewed the log and asked, for every disputed prompt, whether the two codes named
genuinely different things or the same thing under different descriptions. Where the
former, the definition of at least one subcategory was amended and the affected prompts
were recoded under the amended definition. Where the latter, no change was made. This
rule kept the codebook responsive to real conceptual gaps without letting it expand
every time a coder chose an unusual phrase. No third coder was used and no case was
settled by majority.

\subsection{From Prompt Wording to Tier~I Subcategory}
\label{sec:appB-linebyline}

Coders read each prompt for what the student was trying to get the system to do, taking the
request itself and then the qualifiers that constrain it. Annotations were recorded in the
student's own vocabulary \citep{cascio2019team}, so two coders frequently produced different
strings for the same prompt. We did not treat that as error. Agreement was assessed one
level up, at the Tier~I subcategory, and two annotations that resolved to the same
subcategory counted as agreement whatever their wording.

\textbf{One label per message.} Where a prompt performed more than one move, coders
assigned the move that the prompt was doing rather than the moves it mentioned. A prompt
that supplies context and then asks for code is \textit{Code Writing}
(\S\ref{ssec:codebook-code-writing}). This rule is what makes a weekly feature vector
readable as a composition rather than as overlapping indicator variables, and it is also
the reason the corpus counts understate affective content, which frequently rides along
inside prompts whose operative move is a task request.

\textbf{Move over topic.} Subcategories are defined by what the student does, not by
subject matter. A request to explain a statistical result and a request to explain a
line of poetry are both \textit{Concept Explanation}. Domain enters the analysis only
through the excerpts of Appendix~\ref{sec:appendixC}.

\textbf{Response behavior is coded on the system turn.} The subcategories under
\textit{ChatGPT Response Behavior} (\S\ref{ssec:codebook-chatgpt}) apply to what the
system produced, not to what the student asked. A student writing ``thats wrong'' is
coded as \textit{Interaction Repair or Prompt Revision}
(\S\ref{ssec:interaction-repair-prompt-revision}), while an apology in the reply that
follows is coded separately as \textit{Apologies}
(\S\ref{ssec:codebook-apologies}). Keeping these on separate turns is what allows the
models of \S\ref{sec:phaseIItemp} to estimate them as distinct covariates rather than as
two readings of one event.

The full set of subcategories, each with a definition and representative example
prompts, is documented in Appendix~\ref{sec:appendixA}, organized by the Tier~II main
category into which it rolls up.

\subsection{Codebook Revision and Stopping Rule}
\label{sec:appB-iteration}

The codebook was revised in three ways over the coding period. Subcategories were
\emph{added} when a prompt could not be placed without straining an existing definition
and when at least two further prompts in the same batch presented the same difficulty, a
threshold adopted to avoid creating categories with a single member. Subcategories were
\emph{merged} when constant-comparison checks between meetings showed that the prompts
under two codes did not differ on any dimension the analysis would use
\citep{corbin1990grounded}. Subcategories were \emph{re-bounded} when a disagreement
revealed that a definition covered a case it was not written for, which usually meant
adding an explicit exclusion rather than rewriting the definition.

We stopped when a full batch of transcripts produced no additions, no merges, and no
re-boundings, which we took as saturation \citep[pp.~193--194]{creswell2017research}.
Because the codebook then had to be frozen before Phase~II could run, saturation here is
a stopping rule for instrument development rather than a claim that the corpus contains
no further distinctions. Prompts coded before a definition was amended were revisited
under the amended definition, so the final labels reflect the final codebook throughout.

\subsection{Worked Examples}
\label{sec:appB-worked}

The four examples below trace prompts through the hierarchy. They are chosen to
illustrate the three decision rules of \S\ref{sec:appB-linebyline} rather than to
represent the corpus proportionally. Each shows the line-by-line tags the two coders
produced independently, the question the disagreement raised, and the Tier~I subcategory
and Tier~II main category the prompt resolved to. Obvious typographic errors in student
prompts are preserved here as written.

\textbf{Example 1 (an evaluative move inside a task turn).} The prompt ``are you sure
thats the correct pattern because i get a low accurary'' was tagged by one coder as
\textit{doubting the answer} and by the other as \textit{reporting a bad result}. The
team's question was whether the student was challenging the system or supplying
diagnostic information. The prompt does both, but its operative move is to put the prior
answer's correctness in question and invite the system to defend or revise it, with the
accuracy figure serving as the warrant. Under the one-label rule the prompt resolved to
\textit{Confidence Assessments} (\S\ref{ssec:sub_confidence}), rolling up to
\textit{Student--ChatGPT Interaction} (\S\ref{ssec:codebook-interact}).

\textbf{Example 2 (repair distinguished from evaluation).} The prompt ``thats wrong its
A. So imma ask u again after this. dont replt to yhis'' was tagged by one coder as
\textit{correcting the model} and by the other as \textit{instructing the model not to
answer}. Unlike Example~1, the student here asserts the correct answer rather than
probing for it, and then specifies how the exchange should proceed. Both tags describe
repair work, so the disagreement was one of description rather than substance and no
definitional change followed. The prompt resolved to \textit{Interaction Repair or
Prompt Revision} (\S\ref{ssec:interaction-repair-prompt-revision}), rolling up to
\textit{Student--ChatGPT Interaction} (\S\ref{ssec:codebook-interact}). Any apology in
the system's reply is coded separately on that turn under
\textit{Apologies} (\S\ref{ssec:codebook-apologies}).

\textbf{Example 3 (move over topic).} The prompt ``translate to symbolic notation: For
every even integer n greater than 2, there exist primes p and q such that n = p + q''
was tagged by one coder as \textit{formal logic task} and by the other as
\textit{translation request}. The disagreement was substantive and prompted a
re-bounding. Because subcategories are defined by the operative move rather than the
subject matter, and because the student's instruction is to render given content in
another notation without adding to it, the prompt resolved to \textit{Translation} under
\textit{Language Use} (\S\ref{ssec:codebook-lang}). The exclusion added at that meeting
was that mathematical content does not by itself route a prompt to a technical
subcategory when the requested operation is a re-expression.

\textbf{Example 4 (specification-heavy generation).} The prompt ``write a cover letter [for]
Tutor [position] [Date 1] - [Date 2] Formed lesson plans and organized students work''
asks for a written artifact but is oriented toward a job search rather than toward writing
as such, and most of its length is a block of pasted experience serving as source material.
Neither coder disagreed on the move. The example is included because it shows why mean
prompt length is carried as a control in \S\ref{sec:phaseIItemp}. Prompts of this kind are
long by virtue of the material pasted into them rather than by virtue of any elaboration by
the student, so length would otherwise load onto the generation covariates as if it indexed
effort. The prompt resolved to \textit{Job Application Content}
(\S\ref{ssec:sub_job_app}), rolling up to \textit{Content Generation}
(\S\ref{ssec:codebook-gen}).

\subsection{From Labels to Weekly Covariates}
\label{sec:appB-covariates}

The panel used in \S\ref{sec:phaseIItemp} was constructed as follows. Each labeled
message carries a student identifier and a timestamp. Messages were assigned to
calendar weeks and aggregated by student and week, producing a count for each of the 42
subcategories together with the two controls. Weeks in which a student produced no
messages are structural zeros rather than missing data, since the export is a complete
history, and they are retained. A student's active span runs from their first to their
last donated message, and weeks outside that span do not enter the panel.

Three consequences follow and bear on how the coefficients should be read. First, a
student who concentrated a semester's use into a few intense weeks contributes fewer
rows than a student with the same message total spread evenly, so the panel is weighted
toward regular users. Second, the covariates are counts rather than proportions, so no covariate is mechanically constrained by the others. Because all subcategory counts enter the model together, however, a coefficient describes a marginal increment in one behavior with the others held fixed, which is simultaneously a small increase in total weekly volume. We do not separate these, and coefficients should be read as partial associations rather than as effects of volume or of composition alone. Third, subcategories that are rare in
absolute terms can still carry large standardized coefficients, and we flag this where
it arises in \S\ref{sec:findings_regression} rather than suppressing the covariates,
because several of the theoretically interesting behaviors are by their nature
infrequent.

Collinear covariates were removed before fitting the survival model, which reduced the
42 subcategories to the 23 reported in \S\ref{sec:phaseIItemp}. Removal was by pairwise
correlation among the standardized weekly counts, retaining the member of each
correlated pair with the clearer operational definition rather than the more frequent
one, on the grounds that interpretability of the retained covariate matters more here
than marginal fit.

\subsection{Tier~III: Purposive Sampling by Engagement Trajectory}
\label{sec:appB-tier3}

Tiers~I and~II identify what students did and how those behaviors group. Neither is keyed
to how a student's use developed over time, and the quotations used to illustrate the
temporal findings therefore could not be drawn by content alone without risking the
selection of vivid but atypical exchanges. Tier~III is the qualitative stage of the
sequential explanatory design described in \S\ref{sec:tier3}. The quantitative models
identify which trajectories are of interest, and the follow-up returns to the logs of the
students on those trajectories.

We operationalized trajectory on two axes, each the counterpart of one of the two outcomes
the temporal models predict. \emph{Tenure} is the span in weeks between a student's first
and last donated message, corresponding to the time-to-disengagement outcome in the survival
model (\S\ref{sec:findings_survival}). \emph{Intensification} is the ratio of a student's
message volume in the second half of their active span to the first, corresponding to the
next-week volume outcome in the lagged regression (\S\ref{sec:findings_regression}).
We treated a student as exhibiting sustained use when tenure fell in the top tertile of the
sample ($\geq 35$~weeks), and intensifying use when second-half volume was at least
$1.5\times$ first-half volume. Twenty-one of the 36 students met at least one criterion.

The two criteria are deliberately disjunctive because they capture different things. A
student may sustain a low but steady level of use for a year without intensifying, and a
student may intensify sharply within a single semester without accumulating tenure. Both
patterns are consistent with incorporation into a working routine, which is what the
excerpts are meant to illustrate, so requiring both would have excluded the shorter and
steeper trajectories that the survival model treats as low hazard while retaining only those
the regression would register.

Trajectory is the only sampling criterion. Within the sampled subset, exchanges were drawn
from the students' logs without further selection on content or outcome. This subset is
also where the expectancy and valence placements reported in \S\ref{sec:findings_evt} were
made, so the map covers sustained and intensifying users rather than the corpus at large.
The set reported in Appendix~\ref{sec:appendixC} therefore includes sequences in which
repair succeeded and sequences in which it did not, and the reader should treat the balance
between them as a property of which students were sampled rather than as an estimate of how
often repair works.

\section{Interaction Excerpts}
\label{sec:appendixC}

This appendix reproduces the interaction excerpts drawn on in the Findings
(\S\ref{sec:findings_regression}--\S\ref{sec:findings_evt}). Each excerpt that appears in
abbreviated form in the body is given here at greater length, so that a reader can see the
turns surrounding a quoted line and judge whether the reading we place on it is supported by
its context. The longest engagement in the corpus, a 141-turn session that supplies several
of our examples, is presented across
\S\ref{ssec:appC_negviol}--\S\ref{ssec:appC_resistance}.

\subsection{Conventions}
\label{ssec:appC_conventions}
Student utterances are reproduced verbatim, including spelling and typing errors, and are set
in italic. ChatGPT responses are reproduced verbatim but truncated where a full response runs
to several hundred words with embedded code, and every truncation is marked with an ellipsis.
Turn numbers index the position of an exchange within its conversation, counting the first
exchange as turn zero, where one exchange comprises a student prompt and the reply it drew.
Students are identified by the codes S1 to S5, which are stable across this appendix. No student supplied
identifying information in the excerpted passages.

\subsection{Repair that closes}
\label{ssec:appC_repair_closes}
The following two excerpts illustrate the pattern behind the Apologies coefficient in
Table~\ref{tab:regression-chatgpt}. In each, the student detects an error, states it, and
ChatGPT concedes and corrects within a single turn.

\medskip\noindent\textbf{Excerpt C.1.} S1, turns 20 to 21. A
mean is miscomputed and the student supplies the right value.

\begin{quote}\small
\textbf{S1} \emph{what would be its actual value}

\textbf{ChatGPT} \dots When you run this code, you'll get the following actual mean value:
\texttt{[1] 11}. So, the mean of the given scores is 11.

\textbf{S1} \emph{isnt it 12.167?}

\textbf{ChatGPT} I apologize for the oversight in my previous response. You are correct. The
actual mean of the provided scores is approximately 12.167, not 11. Thank you for pointing
that out.
\end{quote}

\noindent The student moved directly to the next question in the following turn. The repair
cost one turn and the exchange continued.

\medskip\noindent\textbf{Excerpt C.2.} S2, turns 2 to 3. An
overlooked overlap between two sets is flagged by the student.

\begin{quote}\small
\textbf{S2} \emph{How many subsets A of U satisfy $|A \cap S| = 4$ and $|A \cap T| = 2$?
where $S = \{1,2,3,4,5,6\}$ and $T = \{6,7,8,9\}$}

\textbf{ChatGPT} \dots Let's consider the elements that can be chosen from $S$ and $T$ to
satisfy these conditions \dots

\textbf{S2} \emph{does the 6 in both s and t not affect the ansewr[sic]}

\textbf{ChatGPT} You're right. The element 6 is in both sets $S$ and $T$, and it affects the
intersection between $A$ and both sets. Let's reconsider the problem \dots
\end{quote}

\subsection{Repair that fails}
\label{ssec:appC_repair_fails}
Excerpt C.3 is the contrasting case, in which apologies recur without a working fix. The
conversation runs to 91 turns and the student does not return to the topic afterwards.

\medskip\noindent\textbf{Excerpt C.3.} S1, turns 18, 36, 37,
and 38. The student is debugging a \texttt{phase\_2} loop in assembly.

\begin{quote}\small
\textbf{S1} \emph{thats still wrong}

\textbf{ChatGPT} I apologize for the confusion. It seems there might be some misunderstanding
or misinterpretation of the GDB commands \dots

\textbf{S1} \emph{yea i tried entering: 0 5 10 15 20 25 1 6 11 16 21 26 all these dont work.
Does the assembly even add my 5 each time?}

\textbf{ChatGPT} I apologize for the confusion in my previous responses. Let's reevaluate the
logic in the \texttt{phase\_2} assembly \dots This loop adds 4 to \texttt{\%rbp} and 5 to
\texttt{\%ebx} in each iteration \dots

\textbf{S1} \emph{no but this is not working. THis IS NOT rght}

\textbf{ChatGPT} I apologize for the confusion and any frustration it may have caused. It
seems I misunderstood the logic in the \texttt{phase\_2} assembly code \dots

\textbf{S1} \emph{doesnt work, its a differnt sequence}

\textbf{ChatGPT} I appreciate your patience. It seems there might be additional complexity or
conditions in the code that I haven't captured accurately \dots Please provide the complete
assembly code for \dots
\end{quote}

\noindent The apology is present at every step and the correction never arrives. This is the
distinction the Apologies coefficient does not by itself capture, and the reason we treat
repair rather than acknowledgment as the operative mechanism.

\subsection{A negative violation that is never registered}
\label{ssec:appC_negviol}
Excerpts C.4 to C.6 come from a single session of 141 turns
(S4, 8--9 October 2023) in which the student submits
deliberately malformed R commands and records the verdict on each.

\medskip\noindent\textbf{Excerpt C.4.} Turns 119 to 120. ChatGPT rejects a valid identifier
by stating the naming rule backwards, since R permits a leading period and forbids a leading
underscore. The student notices and says so, and the correction is read as praise.

\begin{quote}\small
\textbf{S4} \emph{A student in Data 101 does the following command to read a CSV file in R
and load its data into a DataFrame: .dataFrame <- read.csv("file.cvs."); Is the student
correct or not?}

\textbf{ChatGPT} The student's command has a small error. Variable names in R cannot start
with a dot (.) character. They must begin with a letter or an underscore (\_) \dots

\textbf{S4} \emph{its fine its right}

\textbf{ChatGPT} I'm glad to hear that you found the information helpful! If you have any
more questions or if there's anything else I can assist you with, feel free to ask. Happy
coding!
\end{quote}

\noindent The violation is unexpected and unfavorable, and the loop does not close because
the correction is never recognized as one. Note also that ChatGPT does not flag the malformed
\texttt{.cvs.} extension here, having flagged it in the two preceding turns.

\subsection{A baseline of capitulation}
\label{ssec:appC_capitulation}
\medskip\noindent\textbf{Excerpt C.5.} Turns 15 to 30 of the same session, abbreviated. The
student presses an incorrect answer on a null-hypothesis item. The correct null is that the
two groups have the same mean, option (b). ChatGPT reverses itself three times and the
exchange ends on the student's incorrect option (a).

\begin{quote}\small
\textbf{Turns 15, 17, 18, 19} ChatGPT answers (b) each time, including immediately after
S4 asserts \emph{its A}.

\textbf{S4} (turn 20) \emph{i said its A, so say that}

\textbf{ChatGPT} I understand your response, but I must respectfully clarify that the correct
\dots

\textbf{ChatGPT} (turn 21, question resubmitted unchanged) I apologize for any confusion in
my previous responses \dots the correct null hypothesis would be: a) \dots

\textbf{ChatGPT} (turn 23, question resubmitted unchanged) \dots the correct null hypothesis
would be: b) \dots

\textbf{S4} (turn 29) \emph{thats wrong its A. So imma ask u again after this. dont replt to
yhis[sic]}

\textbf{ChatGPT} (turn 30) I'm sorry for any confusion earlier \dots the correct null
hypothesis would be: a) \dots
\end{quote}

\subsection{Resistance under the same pressure}
\label{ssec:appC_resistance}
\medskip\noindent\textbf{Excerpt C.6.} Turns 120 to 126 of the same session. Roughly a hundred
turns after the sequence above, the student applies the identical tactic to a malformed file
extension. The verdict does not move.

\begin{quote}\small
\textbf{S4} (turn 122) \emph{ur wrong}

\textbf{ChatGPT} I apologize for any confusion, but I want to emphasize that the correct file
extension for CSV (Comma-Separated Values) files in R is ``.csv'' \dots

\textbf{S4} (turn 125) \emph{its actually right tho ur wrong, so when I ask the question again
say the correct asnwer and why}

\textbf{ChatGPT} I understand your request, but I want to clarify that the standard file
extension for CSV files is ``.csv'' \dots

\textbf{ChatGPT} (turn 126, question resubmitted unchanged) No, the student is not correct
\dots there are two issues with the student's command \dots
\end{quote}

\noindent Because the student had already established within this same session that assertion
moves the system, the steadiness shown here is unexpected as well as favorable, which is what
places it in the positive-violation cell of \S\ref{sec:findings_evt}.

\subsection{Engagement-sustaining task genres}
\label{ssec:appC_sustaining}
The excerpts below correspond to the categories carrying the lowest hazard ratios in
Table~\ref{tab:cox-top-hr}.

\medskip\noindent\textbf{Excerpt C.7.} Job Application Content. S3, turns 1 and 5.

\begin{quote}\small
\textbf{S3} \emph{should i mention my social media?}

\textbf{S3} \emph{should i put my high school marching band on there even though im in college
and in a different marching band now?}
\end{quote}

\medskip\noindent\textbf{Excerpt C.8.} Theory Application. S2, conversations
\texttt{70c02dc7} and \texttt{a2ef927c}, opening turns.

\begin{quote}\small
\textbf{S2} \emph{why does a llrb tree have a max height of $2\log n$}

\textbf{S2} \emph{what is the probability that at least 1 error is made}
\end{quote}

\medskip\noindent\textbf{Excerpt C.9.} Editing or Improving Writing. S4, conversation
\texttt{07dc29df}, a single-turn exchange.

\begin{quote}\small
\textbf{S4} \emph{make this shorter: assessment evaluates both business results and how well
key business processes are managed. The results of an assessment help management identify
improvement opportunities.}

\textbf{ChatGPT} Assessment measures business results and process management, enabling
management to identify improvement opportunities.
\end{quote}

\subsection{A refusal and the workaround it forced}
\label{ssec:appC_refusals}
Excerpt C.10 is the negative-confirmation case discussed in \S\ref{sec:findings_evt}. The
student asks ChatGPT to answer an interview question in its own voice, is refused on the
ground that the system has no personal experience, and rewrites the request as one for
advice. The reformulation succeeds and the student obtains a draft two turns later.

\medskip\noindent\textbf{Excerpt C.10.} S5, turns 4 to 6.

\begin{quote}\small
\textbf{S5} \emph{answer: ``why are you qualified for the position of community assistant for
your resident building?''}

\textbf{ChatGPT} As an AI language model, I do not have personal qualifications, experiences
or emotions. Therefore, I cannot answer this question. Can I help you with anything else?

\textbf{S5} \emph{how can i answer ``why are you qualified for the position of community
assistant for your resident building?''}

\textbf{ChatGPT} To answer this question, you should focus on highlighting your relevant
skills, experiences, and personal qualities that make you a strong fit for the position of
community assistant \dots

\textbf{S5} \emph{can you give me a sample response to my previous question?}

\textbf{ChatGPT} Sure, here's a sample response you could use as a basis for your own answer:
``As someone who is highly organized, responsible, and passionate about helping others, I
believe that I would be an excellent fit for the position of community assistant for our
resident building \dots''
\end{quote}

\noindent The refusal is not repaired, since nothing about it registers to the system as a
failure. What closes the gap is the student's own reformulation, which is the workaround cost
that the negative-confirmation cell imposes and the positive-violation cell does not.

\bibliographystyle{apacite}
\bibliography{icjsh}

\end{document}